\documentclass[usenatbib]{mn2e}

\usepackage{enumitem}
\usepackage{float}
\usepackage{epstopdf}
\usepackage{verbatim}
\usepackage{nicefrac}
\usepackage{hyperref}
\usepackage{url}
\usepackage{amsmath}
\usepackage{aas_macros}
\usepackage{bm}
\usepackage[table,usenames,dvipsnames]{xcolor}
\usepackage{amsmath}
\usepackage{amssymb}
\usepackage{graphicx}
\usepackage{booktabs}
\usepackage{afterpage}

\defcitealias{2013MNRAS.433.2857M}{MW13}
\defcitealias{amon2016}{Amon et al.\ in prep.}
\defcitealias{wolf2016}{Wolf et al.\ 2016}
\defcitealias{amon2017}{Amon et al.\ in prep.}
\defcitealias{joudaki2017}{Joudaki et al.\ in prep.}

\makeatletter
\newcommand*{\rom}[1]{\expandafter\@slowromancap\romannumeral #1@}
\makeatother

\newcommand{\bfA}{{\mathbfss{A}}}

\usepackage{listliketab}

\author [~Johnson et al.]
{Andrew Johnson$^{1,2}$\thanks{email:
    \href{mailto:asjohnson@swin.edu.au}{\protect\nolinkurl{andrew.johnson.melb@gmail.com}}},
  Chris Blake$^{1,2}$\thanks{email:
    \href{mailto:cblake@swin.edu.au}{\protect\nolinkurl{cblake@swin.edu.au}}},
  Alexandra Amon$^3$, Thomas Erben$^4$, \newauthor Karl
  Glazebrook$^1$, Joachim Harnois-Deraps$^3$, Catherine Heymans$^3$,
  \newauthor Hendrik Hildebrandt$^4$, Shahab Joudaki$^{1,2}$, Dominik
  Klaes$^4$, Konrad Kuijken$^5$, \newauthor Chris Lidman$^6$, Felipe
  A.\ Marin$^{1,2}$, John McFarland$^7$, Christopher B.\ Morrison$^{4,8}$,
  \newauthor David Parkinson$^9$, Gregory B.\ Poole$^{10}$, Mario
  Radovich$^{11}$, Christian Wolf$^{12}$ \\\\
$^1$ Centre for Astrophysics \& Supercomputing, Swinburne University of Technology, P.O.\ Box 218, Hawthorn, VIC 3122, Australia \\ 
$^2$ ARC Centre of Excellence for All-sky Astrophysics (CAASTRO) \\ 
$^3$ Scottish Universities Physics Alliance, Institute for Astronomy, University of Edinburgh, Royal Observatory, Blackford Hill, \\ Edinburgh, EH9 3HJ, U.K. \\ 
$^4$ Argelander Institute for Astronomy, University of Bonn, Auf dem Hugel 71, 53121, Bonn, Germany \\
$^5$ Leiden Observatory, Leiden University, Niels Bohrweg 2, 2333 CA Leiden, The Netherlands\\ 
$^6$ Australian Astronomical Observatory, North Ryde, NSW 2113, Australia \\
$^7$ Kapteyn Astronomical Institute, Postbus 800, NL-9700 AV Groningen, The Netherlands \\
$^8$ Department of Astronomy, University of Washington, Box 351580, Seattle, WA 98195, U.S. \\
$^9$ School of Mathematics and Physics, University of Queensland, QLD 4072, Australia \\
$^{10}$ School of Physics, University of Melbourne, Parkville, VIC 3010, Australia \\
$^{11}$ INAF - Padova Astronomical Observatory, Italy \\ 
$^{12}$ Research School of Astronomy and Astrophysics, The Australian National University, Canberra, ACT 2611, Australia}

\date{\today}

\pubyear{2016}

\def\beq{\begin{equation}} \def\eeq{\end{equation}}

\hypersetup{
    colorlinks=true,
    linkcolor=red,
    citecolor=blue,
} 

\title[Redshift distributions from cross-correlations]{2dFLenS and
  KiDS: Determining source redshift distributions with
  cross-correlations}

\begin{document}
\maketitle

\begin{abstract}
We develop a statistical estimator to infer the redshift probability
distribution of a photometric sample of galaxies from its angular
cross-correlation in redshift bins with an overlapping spectroscopic
sample. This estimator is a minimum-variance weighted quadratic
function of the data: a {\it quadratic estimator}.  This extends and
modifies the methodology presented by \citet{2013MNRAS.433.2857M}.
The derived source redshift distribution is degenerate with the source
galaxy bias, which must be constrained via additional assumptions.  We
apply this estimator to constrain source galaxy redshift distributions
in the Kilo-Degree imaging survey through cross-correlation with the
spectroscopic 2-degree Field Lensing Survey, presenting results first
as a binned step-wise distribution in the range $z < 0.8$, and then
building a continuous distribution using a Gaussian process model.  We
demonstrate the robustness of our methodology using mock catalogues
constructed from N-body simulations, and comparisons with other
techniques for inferring the redshift distribution.
\end{abstract}

\section{Introduction}
\label{sec:intro}

Current and forthcoming photometric surveys aim to image a significant
fraction of the sky\footnote{For example, deep optical imaging surveys
  currently being completed for the science goal of weak gravitational
  lensing include the Kilo-Degree Survey
  \citep{2015A&amp;A...582A..62D}, the Dark Energy Survey
  \citep{2016PhRvD..94b2001A}, and the HyperSuprimeCam imaging survey.
  Future such surveys will include those performed by the Large
  Synoptic Sky Telescope (LSST) and {\it Euclid.}}.  In doing so, they
will obtain the angular positions of millions of galaxies.  Realizing
the scientific potential of these surveys requires an estimate of the
{\it redshift distribution of the galaxies}, which is important for
connecting measurements -- such as tomographic weak lensing
\citep{1999ApJ...522L..21H,2002PhRvD..65f3001H}, the Integrated
Sachs-Wolfe effect and angular power spectra -- to the underlying
cosmological model.  In this work we investigate a method to measure
galaxy redshift distributions using angular cross-correlations with an
overlapping spectroscopic sample.

We outline the approach as follows.  Consider two galaxy datasets: a
spectroscopic sample with a known redshift distribution, and a
photometric sample with an unknown redshift distribution.  The samples
overlap on the sky and in redshift.  Since they are sampled from the
same underlying density field, we expect that they will share a
positive cross-correlation function regardless of galaxy attributes
such as colour and luminosity.  The amplitude of the angular
cross-correlation will increase with the degree of overlap of the two
samples
\citep[e.g.,][]{2008PhRvD..78d3519H,2009A&amp;A...493.1197E,2008ApJ...684...88N}.
Therefore, the redshift distribution of the photometric sample can be
mapped out by dividing the spectroscopic sample into adjacent redshift
bins, and for each bin measuring the angular cross-correlation with
the photometric sample\footnote{As we will discuss below, there are
  additional effects which can correlate the two samples.}.  In this
work we use this technique to constrain the redshift distribution of
galaxies in tomographic bins within the Kilo-Degree Survey (KiDS)
\citep{2015A&amp;A...582A..62D,2015MNRAS.454.3500K} using the
spectroscopic 2-degree Field Lensing Survey
\citep[2dFLenS,][]{2016MNRAS.462.4240B} to trace the surrounding
large-scale structure.

Knowledge of the redshift distribution of the source galaxies is a
critical component of a weak lensing analysis because it is required
to calculate the expected weak lensing signal for a given cosmological
model
\citep[e.g.,][]{2006ApJ...636...21M,2006MNRAS.366..101H,2008MNRAS.389..173K}.
Uncertainty and bias in the source redshift distribution directly
propagates to derived cosmological constraints as one of the most
important astrophysical systematics.  The required level of systematic
error control increases in severity for future surveys: for ``Stage
IV'' dark energy experiments \citep{2013PhR...530...87W} such as the
Large Synoptic Sky Telescope (LSST) and {\it Euclid}, in order to
avoid a degradation of dark energy constraints by more than $50\%$,
the mean and standard deviation of the photometric redshift
distribution need to be measured to an accuracy $\sim 0.002 (1+z)$
\citep{2006MNRAS.366..101H,2015APh....63...81N}.

Many approaches have been proposed for determining source redshift
distributions.  We define {\it direct calibration} methods as those
that calibrate a mapping from the flux in photometric bands to a
galaxy's redshift.  Template-based approaches and machine learning
algorithms both lie in this category\footnote{Examples of machine
  learning algorithms include {\tt SkyNet}
  \citep{2014MNRAS.441.1741G}, {\tt TPZ} \citep{2013MNRAS.432.1483C},
        {\tt ANNz2} \citep{2016PASP..128j4502S} and {\tt MLPQNA}
        \citep{2015MNRAS.452.3100C}.  Examples of template-based
        methods include {\tt BPZ} \citep{2000ApJ...536..571B} and {\tt
          EAZY} \citep{2008ApJ...686.1503B}.}, but there are various
factors that make the above level of accuracy difficult to obtain,
including catastrophic photometric errors, completeness requirements
for spectroscopic training samples, and sample variance
\citep{2010MNRAS.401.1399B,2012MNRAS.423..909C,2014MNRAS.444..129C,2015APh....63...81N}.
We will discuss these factors in the subsequent section.  An
alternative `indirect calibration' approach, which we pursue in this
study, is provided by cross-correlation methods.  In particular, we
focus on the extension and application of the {\it optimal quadratic
  estimation} method proposed by \citet{2013MNRAS.433.2857M}
(henceforth, \citetalias{2013MNRAS.433.2857M}), testing this method
using both simulations and data.

The outline of this paper is as follows.  Sec. (\ref{sec:tech})
introduces the strengths and weaknesses of calibration via
cross-correlations, and highlights the previous work in the field.  In
Sec. (\ref{sec:data}) we introduce the datasets we employ in this study:
the Kilo-Degree Survey, the 2-degree Field Lensing Survey and mock
catalogues built from N-body simulations.  Sec. (\ref{sec:parm})
introduces the background theory, and Sec. (\ref{sec:OQE}) describes the
quadratic estimator we employ to measure the redshift distribution of
galaxies.  We validate our methodology using mock catalogues in
Sec. (\ref{sec:simtests}), and present the results of applying our
methodology to data in Sec. (\ref{sec:results}).  We conclude in
Sec. (\ref{sec:final}).

\section{Strategies for photo-$z$ calibration}
\label{sec:tech}

\subsection{Motivations}

The key point of distinction between direct and indirect calibration
approaches is that the former requires spectroscopic redshifts for a
subsample of the full photometric sample, and this subsample needs to
be representative of the full sample in both colour and magnitude
space \citep{2014MNRAS.445.1482S,2016PASP..128j4502S}. This requires
that the targetted spectroscopic sample should be highly complete,
i.e., a secure redshift needs to be measured for $>90\%$ of the
subsample \citep{2015APh....63...81N}.  To achieve this level of
completeness, spectroscopic redshifts are required for faint and
high-redshift galaxies that are abundant in deep imaging surveys.  In
contrast, for a cross-correlation analysis one is free to target any
tracer of overlapping large-scale structure (most usefully, the
brightest galaxies), circumventing this difficulty.

Achieving a high level of spectroscopic-redshift completeness for
direct calibration methods presents a significant observational
challenge, as the chance of obtaining a successful redshift is
dependent on an object's magnitude.  Therefore, spectra are typically
obtained for a non-random subsample of the target catalogue.  A useful
example is the DEEP2 survey conducted on the DEIMOS spectrograph at
Keck Observatory: for the highest redshift quality class, secure
redshifts were only obtained for $60\%$ of the galaxies
\citep{2013ApJS..208....5N}.  Considering future surveys, the severity
of this problem is demonstrated by the requirement for spectroscopic
follow-up suggested by \citet{2015APh....63...81N}: obtaining $>90\%$
redshift completeness at $i=25.3$ (LSST depth) would require more than
100 nights on Keck.

We note that the requirements for spectroscopic follow-up can be
reduced by assigning weights to galaxies during the training phase of
photometric-redshift calibration
\citep{2008MNRAS.390..118L,2012MNRAS.423..909C,2016PASP..128j4502S}. These weights are
assigned based on the colour-magnitude phase-space distribution of
both the parent photo-$z$ sample and the follow-up spectroscopic
sample.  The result is that the weighted spec-$z$ sample more closely
matches the photo-$z$ sample in colour-magnitude space.  The extent to
which this approach allows one to reduce the required completeness is
currently a subject of study.

The above challenges are also relevant for machine learning
algorithms, as they require separate training, testing and validation
samples.  Similar requirements exist for template-based approaches,
where -- even though templates can be derived synthetically in
principle -- assessing the resulting accuracy of the photo-$z$
estimate requires a representative spec-$z$ sample, which can also aid
in the construction of accurate spectral templates.  Moreover,
deriving a Bayesian prior used in the fitting process \citep[e.g.,
  {\tt BPZ}][]{2000ApJ...536..571B} requires a spec-$z$ sample; this
prior can strongly influence the final results
\citep{2014MNRAS.445.1482S}.

{\it Catastrophic errors}, photo-$z$ estimates $z_p$ with $|z_p -
z_{\rm true}| \sim \mathcal{O}(1)$, present an additional issue for
direct calibration methods.  Such errors occur because, with only
broad-band flux information, there exist degeneracies in galaxy
colours such as the confusion between the Lyman and Balmer breaks.
General studies of the consequences of catastrophic errors are
presented by \citet{2010MNRAS.401.1399B} and
\citet{2010ApJ...720.1351H}.  These outliers cannot be mitigated by
re-weighting the sample.

\subsection{Challenges}

Although photo-$z$ calibration by cross-correlation avoids some of the
issues listed above, there remain significant challenges for this
approach, some of which we outline below.

\begin{itemize}

\item{{\bf Degeneracy with galaxy bias}: Cross-correlations measure
  the combination $b(z) \times P(z)$, where $b(z)$ is the source
  galaxy bias factor and $P(z)$ is the source redshift probability
  distribution.  Therefore, galaxy bias is degenerate with the
  redshift distribution \citep{2008ApJ...684...88N}.  Calibrating the
  redshift-dependent galaxy bias requires extra probes (e.g.,
  galaxy-galaxy lensing) or assumptions (e.g., the bias varies
  smoothly with redshift).  This likely represents the dominant issue
  when constraining the source redshift distribution using
  cross-correlations.}  \vspace{0.1cm}
\item{{\bf Cosmological dependence}: The model cross-correlation
  function depends on both our guess of $P(z)$ and the cosmological
  model. This introduces a worrying circularity, as our aim is to test
  the cosmological model with measurements derived using
  $P(z)$.}  \vspace{0.1cm}
\item{{\bf Extra source of correlations}: Cosmic magnification
  introduces additional correlations and hence can bias measurements
  of the redshift distribution.  In addition to changing the shape of
  galaxies, lensing changes their size and brightness, promoting
  fainter galaxies into a magnitude-limited survey and correlating
  foreground and background objects
  \citep{2010MNRAS.401.1399B,2013MNRAS.433.2857M,2014MNRAS.437.2471D}.}
\item{{\bf Spec-$z$ coverage}: The redshift range over which the
  source distribution can be reconstructed is limited by the redshift
  and areal coverage of the spectroscopic cross-correlation samples.}

\end{itemize}

\subsection{Developments}
\label{status}

In this section we summarize recent work on photometric calibration
with angular cross-correlations.

\begin{itemize}

\item{{\bf Estimators:} A number of estimators have been proposed for
  inferring the redshift distribution of a photometric sample using an
  overlapping spec-$z$ sample
  \citep{2008ApJ...684...88N,2010ApJ...721..456M,2010ApJ...724.1305S,
    2013arXiv1303.4722M, 2013MNRAS.431.3307S, 2013MNRAS.433.2857M}.
  For example \citet{2013MNRAS.433.2857M} develop a quadratic
  estimator, while \citet{2010ApJ...724.1305S},
  \citet{2013arXiv1303.4722M} and \citet{2013MNRAS.431.3307S} use
  maximum-likelihood approaches to infer the source distribution.}
\vspace{0.1cm}
\item{{\bf Self-calibration:} Dividing a photometric sample into
  redshift bins allows one to cross-correlate between bins. This
  correlation allows one to determine the contamination fraction for
  the sample, and potentially constrain other systematic errors
  \citep{2007MNRAS.378..852P,2009A&amp;A...493.1197E,2013MNRAS.431.1547B,2016MNRAS.463.3737C}.}
\vspace{0.1cm}
\item{{\bf Applications:} The methodology presented by
  \citet{2013arXiv1303.4722M} and \citet{2013MNRAS.431.3307S} has been
  applied to estimate the redshift distributions of galaxies in the
  Sloan Digital Sky Survey (SDSS)
  \citep{2015MNRAS.447.3500R,2016MNRAS.457.3912R}, the Cosmic Infrared
  Background \citep{2015MNRAS.446.2696S}, the Canada-France-Hawaii
  Telescope Legacy Survey (CFHTLS) \citep{2016MNRAS.462.1683S},
  infrared sources from Wide-field Infrared Survey Explorer (WISE) and
  the Two-Micron All-Sky Survey (2MASS), and radio sources from the
  Faint Images of the Radio Sky at Twenty cm (FIRST) survey
  \citep{2013arXiv1303.4722M}.  Additionally,
  \citet{2016arXiv160605338H} present the first application of this
  methodology to a cosmic shear analysis.}
\end{itemize}

\section{Datasets}
\label{sec:data}

\begin{figure*}
\centering
\includegraphics[width=15.0cm]{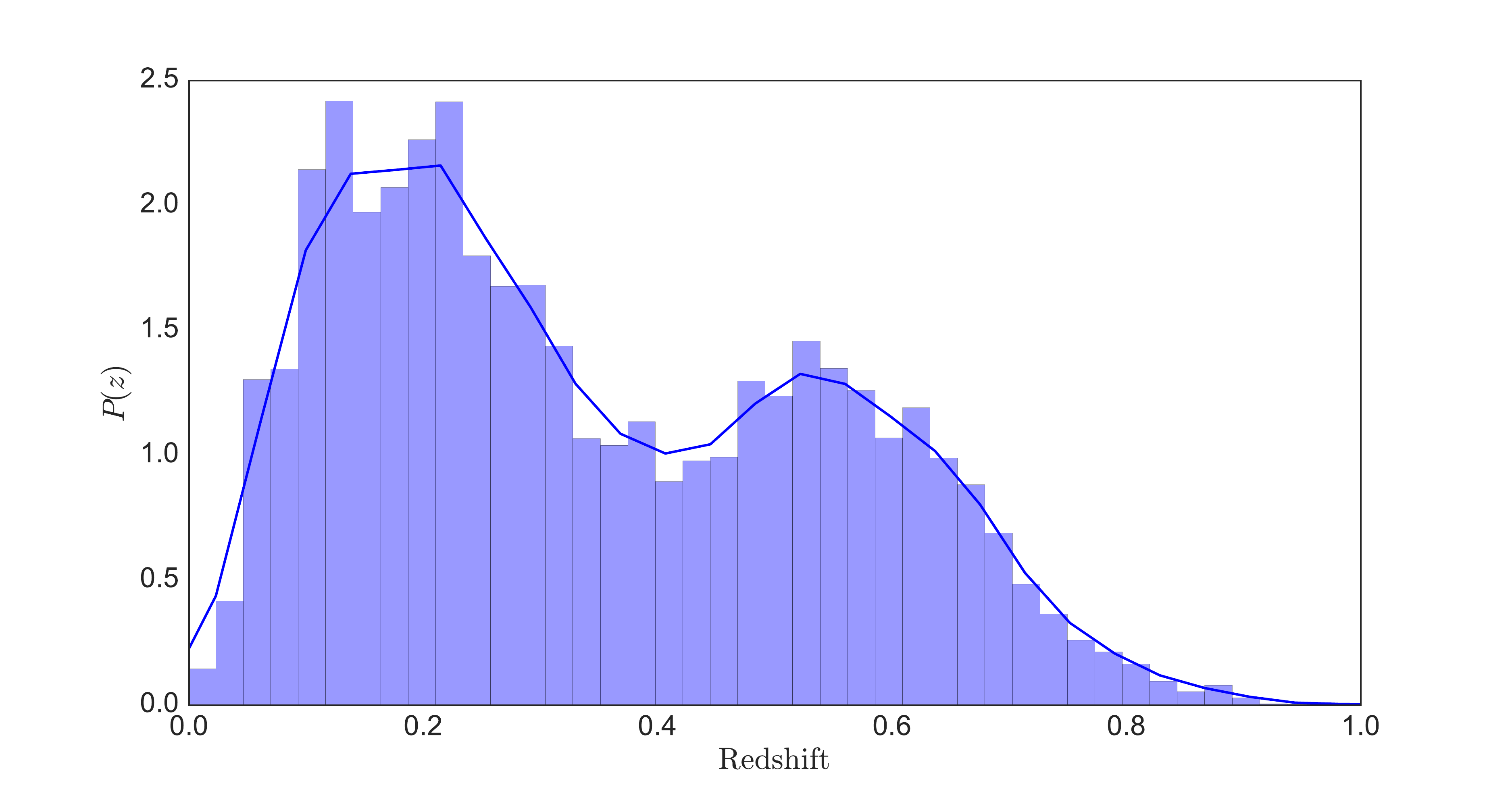}
\caption{The redshift probability distribution of 2dFLenS galaxies,
  displayed as both a histogram and Gaussian kernel density estimate.
  The multiple peaks arise because separate 2dFLenS LRG samples with
  different colour and magnitude selection criteria have been merged.}
\label{nz}
\end{figure*}

\begin{figure*}
\centering
\includegraphics[width=15.0cm]{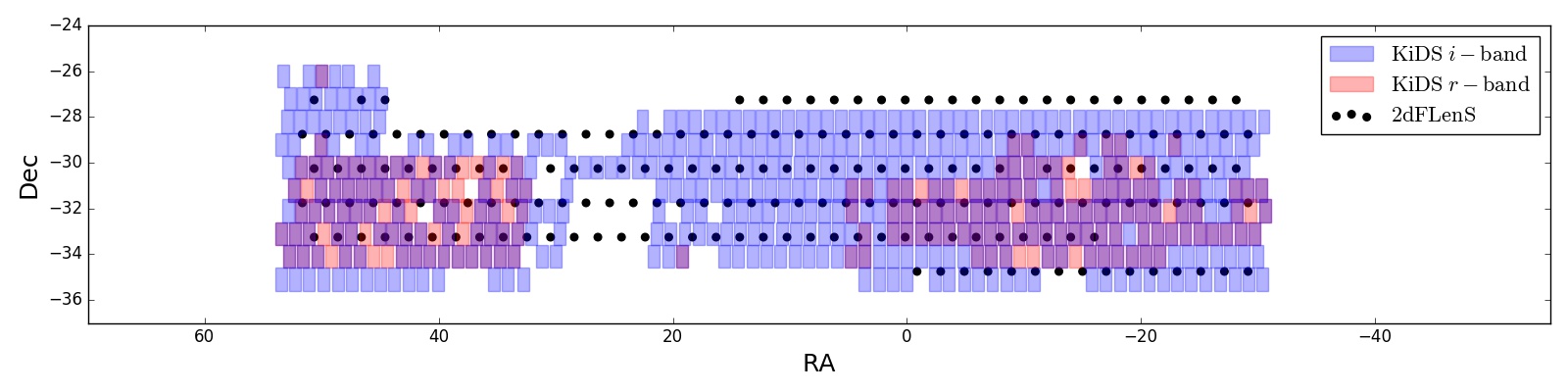}
\caption{The overlap area between 2dFLenS and KiDS.  2dFLenS pointings
  are displayed as black dots (which are centres of circular fields of
  radius 1 deg), and the KiDS-800 $i$-band (KiDS-450 $r$-band)
  coverage is shown as blue (red) coloured tiles.  The total area of
  overlap between 2dFLenS and the $i$-band ($r$-band) imaging is $431$
  deg$^2$ ($152$ deg$^2$).  Each KiDS pointing has dimension $1 {\rm deg}
  \times 1 {\rm deg}$.}
\label{skyplot}
\end{figure*}

\subsection{The Kilo-Degree Survey}
\label{sec:kids}

The Kilo-Degree Survey (KiDS) is a multi-band imaging survey designed
for weak gravitational lensing analyses
\citep{2015A&amp;A...582A..62D}.  The survey is being performed at the
2.6-metre VLT Survey Telescope where, using the $300$-mega-pixel
wide-field camera OmegaCAM, images are taken in four filters
$ugri$. KiDS aims to image $\sim 1500$ square degrees of the sky down
to a limiting $r$-band magnitude of $\sim 25$\footnote{The $r$-band
  images are used for galaxy shape measurements because these are the
  deepest observations obtained in the best seeing conditions.  The
  $ugri$ bands have $5\sigma$ limiting magnitudes $\sim 24.3, 25.1,
  24.9, 23.8$, respectively, in a $2^{\arcsec}$ aperture
  \citep{2016arXiv160605338H}.}.

The first and second data releases of KiDS are presented by
\citet{2015A&amp;A...582A..62D} and \citet{2015MNRAS.454.3500K}. Based
on these catalogues, gravitational lensing science analyses of the
$100$ square degree overlap area with the GAMA survey
\citep{2011MNRAS.413..971D} were undertaken by
\citet{2015MNRAS.452.3529V}, \citet{2015MNRAS.454.3938S},
\citet{2016MNRAS.459.3251V} and \citet{2016MNRAS.462.4451B}, using
matched-aperture $ugri$ colours in conjunction with {\tt BPZ} to
derive redshift probability distributions and hence the lensing
efficiencies.

We performed our analyses using the third data release of KiDS,
separately using the $r$-band ``KiDS-450'' \citep{2016arXiv160605338H}
and $i$-band ``KiDS-800'' \citepalias{amon2016} imaging datasets.
\citet{2016arXiv160605338H} carried out a careful analysis of
different methods for calibrating the source redshift distribution and
settled on a direct photo-$z$ calibration scheme, verified by
estimates based on clustering in a few square degrees of overlapping
deep spectroscopic fields.  \citet{2016arXiv160909085M} have recently
presented a determination of the KiDS redshift distribution using
small-scale cross-correlations.

\subsection{The 2-degree Field Lensing Survey}
\label{sec:2df}

We map the large-scale structure in which our photometric sample is
embedded using the spectroscopic 2-degree Field Lensing Survey.  We
outline the basic properties of the survey here; full details are
presented by \citet{2016MNRAS.462.4240B}.

The principal aim of 2dFLenS is to expand the area of overlap between
spectroscopic galaxy surveys and gravitational lensing imaging
surveys.  This facilitates two key science goals.  First, it allows a
joint analysis of lensing and galaxy redshift samples including all
cross-correlation statistics
\citep[e.g.,][]{2012MNRAS.422.2904G,2012MNRAS.422.1045C}, with
different applications presented by \citetalias{amon2017} and
\citetalias{joudaki2017}.  Second, it allows the calibration of
photometric-redshift distributions using cross-correlation techniques
-- which we present in this paper -- and direct calibration techniques
\citepalias{wolf2016}.

2dFLenS was conducted on the Anglo-Australian Telescope over 53 nights
in the 14B, 15A and 15B semesters.  The two main target classes,
selected from the VST-ATLAS Survey \citep{2015MNRAS.451.4238S},
comprised $\sim 40{,}000$ Luminous Red Galaxies (LRGs) across a range
of redshifts $z < 0.9$ selected by SDSS-inspired cuts
\citep{2013AJ....145...10D}, and a magnitude-limited complete sample
of $\sim 30{,}000$ objects in the range $17 < r < 19.5$ to assist with
direct photometric calibration of the SkyMapper Survey
\citepalias{wolf2016}.  In our study we analyze the LRG sample, whose
normalized redshift distribution (after merging the multiple target
classes) is illustrated in Fig. (\ref{nz}).

2dFLenS observations cover an area of $731$ deg$^2$.  The overlap area
with the imaging is currently limited by the progress of KiDS, which
is still collecting data at the time of writing.  Currently, the
overlap area between 2dFLenS and the $i$-band ($r$-band) KiDS imaging
is $431$ deg$^2$ ($152$ deg$^2$), as shown in Fig. (\ref{skyplot}).

We map out the photo-$z$ redshift distribution through
cross-correlations by dividing our spectroscopic sample into
independent redshift bins.  In choosing the width of these bins we
balanced considerations of noise in the measurements with the desire
for high redshift resolution, defining 18 redshift bins of width
$\Delta z = 0.05$ in the redshift range $0 < z < 0.9$.

\subsection{Simulations}
\label{sec:simulations}

We tested the robustness of our methodology by constructing {\it
  synthetic galaxy catalogues} composed of overlapping photometric and
spectroscopic samples, which allowed us to compare the redshift
distributions reconstructed by our algorithm to the known input
distributions.  We generated these mock catalogues using the Scinet
LIght Cone Simulations (SLICS) series of N-body simulations
\citep{2015MNRAS.450.2857H} which have been produced using the
CUBEP$^3$M code \citep{2013MNRAS.436..540H} using a WMAP9+BAO+SN
cosmological parameter set: matter density $\Omega_m = 0.2905$, baryon
density $\Omega_b = 0.0473$, Hubble parameter $h = 0.6898$, spectral
index $n_s = 0.969$, and normalization $\sigma_8 = 0.826$.  The
box-size of the simulations is $L = 505 \, h^{-1}$ Mpc.  The
simulations follow the non-linear evolution of $1536^3$ particles
inside a $3072^3$ grid cube.  For each simulation, the density field
is output at 18 redshift snapshots in the range $0 < z < 3$, which are
used to construct a survey cone spanning 60 deg$^2$.  A spherical
overdensity halo finder was executed on the particle data during the
simulation run, and the resulting halo catalogues were post-processed
to self-consistently sample the light-cone geometry.  We sampled our
mocks from these halo catalogues, as described further in
Sec. (\ref{sec:simtests}).

\section{Parameterization and Modelling}
\label{sec:parm}

In this section, we describe the redshift distribution
parameterization and clustering model we adopt for our analysis.
Throughout, we will assume the fiducial cosmological parameters of the
SLICS simulations, stated above.  For readability, we will begin with
a heuristic description and then move on to a more rigorous treatment.

The intent of this work is to present a novel technique for measuring
the redshift probability distribution of a given photometric sample of
galaxies, $P^{(p)}(z)$.  First, we need a method to parameterise this
probability distribution.  We do this by dividing the redshift range
of the sample into step-wise bins and constraining the number of
galaxies $N^{(p)}_i$ within each bin $i$, assuming that their
probability distribution within the bin is constant.  So if $W_i(z)$
is a normalized top hat filter ($\int W_i(z) \, dz = 1$) and
$N^{(p)}_{\rm T} = \sum_i N^{(p)}_i$ is the total number of
photometric galaxies, then the probability distribution within each
bin is $P^{(p)}_i(z) = W_i(z)$ and the total distribution is
\begin{equation}
P^{(p)}(z) = \sum_i (N^{(p)}_i/N^{(p)}_{\rm T}) \, W_i(z) \, ,
\label{eq:P}
\end{equation}
which is normalized such that $\int P^{(p)}(z) \, dz = 1$.

The quantity of interest for constraining $P^{(p)}(z)$ is the
cross-power spectrum between two samples of galaxies: specifically,
between the full photometric sample $(p)$ and a given redshift bin of
the spectroscopic sample $(s_i)$.  We label this angular galaxy
cross-power spectrum $C^{(g)}_{p s_i}(\ell)$, as a function of
multipole $\ell$.  One can estimate this quantity by decomposing the
projected density field for each sample into spherical harmonic
coefficients $\{ p(\ell)$, $s_i(\ell) \}$ for the photometric and
spectroscopic sample, respectively. The statistical estimate is then
${\widehat C}_{p s_i}(\ell) = \langle p(\ell) \, s_i(\ell) \rangle$.
In the remainder of this section, we will describe how we can model
this quantity using our parameterization for $P^{(p)}(z)$, and convert
this model from a power spectrum to the measured correlation function.

Using the small-angle (or `Limber') approximation
\citep{1954ApJ...119..655L}, the matter cross-power spectrum between
two redshift bins $i$ and $j$, with redshift distributions $P_i(z)$, is
\begin{equation}
C^{(m)}_{i j}(\ell) = \int_0^\infty dz \, P_i(z) \, P_j(z) \,
\frac{{\mathcal P}(k,z)}{r^2(z) r_{\rm H}(z)} \, ,
\label{eq:calP}
\end{equation}
where ${\mathcal P}(k,z)$ is the matter power spectrum at wavenumber
$k$, $r(z)$ is the co-moving distance to redshift $z$, $r_{\rm H}(z)
\equiv dr/dz$ and $k = (\ell + 1/2)/r$.

To extend Eq. (\ref{eq:calP}) to model the galaxy-galaxy power
spectrum we need to model the galaxy biases for both samples -- which
we label $b^{(p)}(z)$, $b^{(s)}(z)$ -- and to include a shot noise
component.  We will assume a linear relationship between the galaxy
and matter density field, i.e., $\delta_g = b \, \delta_m$
\footnote{This assumption requires explanation.  First, the scales of
  relevance for the quadratic estimator are $\sim 10$ arcmin.  This
  angular scale corresponds to a set of physical scales where we might
  expect linear galaxy bias to break down and introduce a systematic
  error.  However, as emphasized by \citetalias{2013MNRAS.433.2857M},
  the smoothness of the weighting function implies that the Fourier
  modes being traced are on more linear scales than expected from this
  simplistic conversion.  The mock catalogues provide a way for us to
  quantify the significance of this error.  For future work, this
  investigation could be extended by up-weighting linear scales
  (reducing the overall constraining power) or introducing a
  non-linear galaxy bias component (at the cost of further
  complicating the methodology).}, and model the galaxy bias in
step-wise bins as
\begin{equation}
b^{(A)}(z) = b^{(A)}_i
   ~~~\textrm{ for $|z-z_i|<\Delta_i/2$} \\ .
 \label{eqn:bias}
\end{equation}
$(A)$ can be either $(p)$ or $(s)$, indicating the photometric or
spectroscopic sample, respectively, and $\Delta_i$ is the width of the
$i^{\rm th}$ redshift bin. The galaxy cross-power spectrum can then be
written as
\begin{equation}
\begin{split}
C^{(g)}_{p s_i}(\ell) &= \int_0^\infty dz \, P^{(p)}(z) \, b^{(p)}(z)
\, P^{(s)}_i(z) \, b^{(s)}_i(z) \, \frac{{\mathcal P}(k,z)}{r^2(z)
  r_{\rm H}(z)} \\ &+ \omega_{p s_i} \, ,
\end{split}
\label{eq:calP2}
\end{equation}
where $P^{(s)}_i(z)$ is the probability distribution of the
spectroscopic sample in the $i^{\rm th}$ bin and $\omega_{p s_i}$
models the shot noise component.

We now wish to expand this expression in terms of $N^{(p)}_i$.
Applying our parameterisations for $P^{(p)}(z)$, $P^{(s)}_i(z)$,
$b^{(p)}(z)$ and $b^{(s)}_i(z)$ from Eq. (\ref{eq:P}, \ref{eq:calP},
\ref{eqn:bias}), Eq. (\ref{eq:calP2}) reduces to
\begin{equation}
 C^{(g)}_{p s_i}(\ell) = ( N^{(p)}_i /N^{(p)}_{\rm T}) \, b^{(p)}_i \,
 b^{(s)}_i \, C^{(m)}_{i i}(\ell) + \omega_{p s_i} \, .
\label{eqn:Cross1}
\end{equation}
Following a similar derivation, we compute the auto-correlations
between the full photometric sample ($C^{(g)}_{p p}(\ell)$) and
between the bins of the spectroscopic sample ($C^{(g)}_{s_i
  s_i}(\ell)$) as:
\begin{align}
C^{(g)}_{p p}(\ell) & = \sum_i \left( \frac{N^{(p)}_i
  b^{(p)}_i}{N^{(p)}_{\rm T}} \right)^2 C^{(m)}_{i i}(\ell) +
\omega_{p p} \, , \label{CrossAutode2} \\ C^{(g)}_{s_i s_i}(\ell) & =
\left( b^{(s)}_i \right)^2 C^{(m)}_{i i}(\ell) + \omega_{s_i s_i} \,
. \label{CrossAutode3}
\end{align}
Assuming the Limber approximation, which ignores long-wavelength
modes, the covariance between non-overlapping bins is zero: thus,
$C^{(g)}_{s_i s_j}(\ell) = 0$ when $i \ne j$.  For further details see
\citetalias{2013MNRAS.433.2857M}.

To model the shot noise components for
Eq. (\ref{eqn:Cross1},\ref{CrossAutode2},\ref{CrossAutode3}) we assume
Poisson statistics, neglecting non-Poisson contributions
\citep{2013PhRvD..88h3507B}.  Following this assumption the shot noise
components are $\omega_{A_i A_i} = N^{(A)}_i / {\rm area}~ [{\rm
    steradian}]$ and $\omega_{p s_i} = f_{\rm over}~N^{(s)}_i / {\rm
  area}~ [{\rm steradian}]$, where $f_{\rm over}$ is the overlap
fraction between the photo-$z$ and spec-$z$ sample.

For observational considerations we will switch to configuration space
when applying our methodology to data.  Thus, we need to transform
Eq. (\ref{eqn:Cross1},\ref{CrossAutode2},\ref{CrossAutode3}) into
configuration space.  To simplify the final expressions we first
transform the constant number count case of Eq. (\ref{eq:calP}):
\begin{equation}
w^{(m)}_{i i}(\theta) \equiv \sum_\ell \frac{2\ell +1}{4\pi} ~
P_{\ell}(\cos \theta) \, C^{(m)}_{i i}(\ell) \, .
\end{equation}
Now the angular galaxy auto- and cross-correlation functions can be
written as
\begin{align}
w_{p s_i}(\theta) &= ( N^{(p)}_i /N^{(p)}_{\rm T}) ~ b^{(p)}_i ~ b^{(s)}_i ~ w^{(m)}_{i i}(\theta) \, , \label{wtheta1} \\
w_{s_i s_i}(\theta) & =  (b^{(s)}_i)^2 w^{(m)}_{i i}(\theta)  \, , \label{wtheta2} \\
w_{p p}(\theta) & = \sum_i \left( \frac{ N^{(p)}_i b^{(p)}_i}{N^{(p)}_{\rm T}} \right)^2 w^{(m)}_{i i}(\theta) \, . \label{wtheta3}
\end{align}

To compute the various correlation statistics we use the public
software {\tt CHOMP}\footnote{\protect\url{https://github.com/morriscb/chomp}}
introduced by \citet{2013JCAP...11..009M}\footnote{We checked the
  accuracy of this code by comparing its output with our own
  calculations. For both the angular power spectrum and correlation
  function, the calculations agree. We adopt {\tt CHOMP} because of
  its useful class-based structure.}.  This calculation requires as
input the matter power spectrum for each redshift bin, which we model
using the {\tt HaloFit} code \citep{2003MNRAS.341.1311S}: the {\tt
  HaloFit} parameters adopted are those fit by
\citet{2012ApJ...761..152T}.  {\tt CHOMP} computes the redshift
evolution in each bin from linear theory using ${\mathcal P}(k,z) =
D(z)^{2}{\mathcal P}(k)$, where $D(z)$ is the growth function. This
approximation is valid because the redshift bins we adopt are narrow.

We note a number of systematic modelling issues which could be improved in 
future analysis:

\begin{itemize}
\item{{\it Non-linear effects}. We measure the angular correlation
  function to scales $\sim 1$ arcmin.  On such scales non-linear
  effects become significant and the {\tt HaloFit} model we adopt may
  become inaccurate.  We also assume linear galaxy bias.}\vspace{0.1cm}
\item{{\it Bias evolution}. The cross-correlation observable is the
  combination $b^{(p)}(z) P^{(p)}(z)$, such that our inference of
  $P^{(p)}(z)$ must depend on the redshift evolution of galaxy bias.}\vspace{0.1cm}
\item{  {\it Flat $N(z)$ approximation}. To derive the above equations
  we have approximated the redshift distributions using a step-wise
  parametrization, such that the redshift distribution is constant
  within each bin.  This approximation will break down if there are
  steep gradients in the redshift distribution.}
\end{itemize}

\section{The Quadratic Estimator}
\label{sec:OQE}
In this section we outline the construction and properties of the
`quadratic estimator' we employ to measure the redshift distribution
using cross-correlations. This work extends that presented by
\citetalias{2013MNRAS.433.2857M}.

\subsection{Introduction}
\label{OQE:background}

By {\it quadratic estimator} we are referring to a statistical
estimator of a quantity, say $N$, based on a quadratic combination of
the available data, ${\bf x}$. For example, $ \widehat{N} = {\bf x}^T
{\bf x}$.  The symbol ($\widehat{\dots}$) indicates a
statistically-estimated quantity: a value derived directly from data
rather than the true value.  Quadratic estimation is particularly
relevant for Gaussian random fields as all the information content is
contained within quadratic combinations of the data (second-order
statistics).

To construct a quadratic estimator we need to $(i)$ define the data,
$(ii)$ specify the quantity we wish to estimate, and $(iii)$ construct
a method to combine the data that gives an estimate of the desired
quantity -- preferably this estimator will take advantage of all of
the information content within the data, thus minimising the final
variance of the inferred parameter (such an estimator is said to
satisfy the Cramer–-Rao inequality and be {\it optimal}).  Considering
each of these points in turn:

\begin{itemize}
\item \textbf{\emph{The data}}: We start by considering the spherical
  harmonic coefficients of the projected density fields, which we
  write as $\widehat{p}(\ell,m)$ and $\widehat{{\bf s}}(\ell,m)$ for
  the photo-$z$ and spec-$z$ samples, respectively, where ${\bf s}
  \equiv s_i$ represents the coefficients for the $i^{\rm th}$
  spec-$z$ bin.  These coefficients are computed as follows. First, we
  define $n(\Omega)$ as the projected galaxy density field, where
  $\Omega$ indicates the angular position.  The spherical harmonic
  coefficients are computed by projecting the density field onto a
  basis of spherical harmonics $(Y^{m}_{\ell})$:
\begin{equation}
\widehat{p}(\ell,m) = \frac{1}{\bar{n}} \int ~ d\Omega ~ n(\Omega) ~
Y^{m}_{\ell} (\Omega) \, .
\end{equation}
We combine these coefficients into a single data vector ${\bf x} =
\left( \widehat{p}(\ell,m), \widehat{{\bf s}}(\ell,m)\right)$.

\vspace{0.1cm}
\item \textbf{The estimated quantity}: The quantity we wish to
  determine is the number count distribution of the photometric sample
  in step-wise bins, labelled $\widehat{N}^{(p)}_i$ (following the
  parameterization defined above).\vspace{0.1cm}
\item \textbf{The estimator}: We begin by writing the estimator in the
  most general form possible: $\widehat{N}_i = {\bf x}^T {\bf E}_i
  {\bf x} - c_i \,$, where ${\bf E}_i$ is a symmetric matrix and $c_i$
  is a constant
  \citep{1997PhRvD..55.5895T,1998PhRvD..57.2117B,Dodelson-Cosmology-2003}.
  These free parameters will be fixed by imposing various conditions
  on the estimator. Rather than making a single estimate, we can
  iterate until we are satisfied with the convergence. Setting
  $[\widehat{N}_i]_{\rm last}$ as our initial guess, the updated
  estimator is
\begin{equation}
\widehat{N}_i = [\widehat{N}_i]_{\rm last} + {\bf x}^T {\bf E}_i {\bf x} - c_i \, .
\end{equation}
\end{itemize}
Requiring that the estimator is unbiased and optimal one can solve for
both free parameters.  Being {\it unbiased} implies that an ensemble
average of the estimates converges to the input or true value,
$\langle \widehat{N}_i \rangle = N^{\rm true}_i$.  Being {\it optimal}
implies that the estimator minimizes the variance, viz., $\langle
\widehat{N}_i^2 \rangle - \langle \widehat{N}_i \rangle^2$ is
minimised.  The final form of the estimator is
\citep{1997PhRvD..55.5895T}
\begin{equation}
\widehat{N}_i = [\widehat{N}_i]_{\rm last} + \frac{1}{2}
\sum_j[{\bf F}^{-1}]_{ij} \bigg[ {\bf x}^{T}{\bf Q}_j{\bf x} -
  {\rm Tr}({\bf Q}_j {\bf A}) \bigg] \, ,
\label{GeneralEstimator}
\end{equation}
where
\begin{equation}
{\bf Q}_j = {\bf A}^{-1}{\bf A}_{,j}{\bf A}^{-1} \, ,
\end{equation}
${\bf A} \equiv \langle {\bf x} {\bf x}^T \rangle$ is the covariance matrix
of the data, and its derivative is ${\bf A}_{,\alpha} = \partial {\bf
  A}/ \partial {N}_{\alpha}$.  We note that implicitly ${\bf A}$ is a
function of both $\ell$ and $m$, i.e., ${\bf A} = {\bf A}(\ell,m)$.
Assuming many modes are included one can approximate ${\bf F}$ as the
Fisher matrix\footnote{The Fisher matrix is an approximation of the
  curvature matrix, equal to its ensemble average.  For details see
  \citet{1998PhRvD..57.2117B}.}
\begin{equation}
F_{ij} = \frac{1}{2} \sum_{\ell, m}{\rm Tr} \left[\bfA^{-1} \, \bfA_{,i}  \,\bfA^{-1} \bfA_{, j} \right] \, . 
\label{eqn:fish1}
\end{equation}
When this assumption is violated the Fisher matrix will be biased by
sample variance.  We do not expect this assumption to have a
significant effect on our results.

Note, from the previous section
(Eq. \ref{eqn:Cross1},\ref{CrossAutode2},\ref{CrossAutode3}), ${\bf A}$
is known. It is the full covariance matrix between the spec-$z$ and
photo-$z$ samples, including auto-correlations.  In particular,
\begin{equation}
{\bf A} = \langle {\bf x} {\bf x}^T \rangle = \Bigg\langle \left( \widehat{p}(\ell,m) ~~ \widehat{{\bf s}}(\ell,m) \right)
\begin{pmatrix}
         \widehat{p}(\ell,m) \\
         \widehat{{\bf s}}(\ell,m) 
        \end{pmatrix}
\Bigg\rangle \nonumber
\end{equation}
\vspace{-0.2cm}
\begin{equation}
 = \begin{pmatrix}
         C^{(g)}_{p p}(\ell) ~ C^{(g)}_{p {\bf s}}(\ell) \\
         C^{(g)}_{{\bf s} p}(\ell) ~ C^{(g)}_{\bf{s} \bf{s}}(\ell) 
        \end{pmatrix}	
        \, .
        \label{eqn:A}
\end{equation}
Moreover, from Eq. (\ref{eqn:Cross1}), the derivative of the
off-diagonal terms is ${\partial A_{0i}}/{\partial N_i} = b^{(p)}_i
b^{(s)}_i N^{(s)}_i C_{s_i s_i}(\ell)$.

A more intuitive derivation of Eq. (\ref{GeneralEstimator}) can be
found by applying the Newton-Raphson method to the Gaussian likelihood
function of $\bf{x}$, $\mathcal{L(\bf{x})}$. The basic idea is to
solve, iteratively, for roots of $\partial \ln \mathcal{L} / \partial
N_i$: the roots indicate a maximum of the likelihood function.  For
this alternative derivation we refer the reader to
\cite{1998PhRvD..57.2117B}.

\subsection{Revised form of the estimator}
\label{OQE:corrections}

In this section and the subsequent one, we present important yet
rather tedious mathematics; therefore, in the aid of readability we
offer a quick summary.

The primary purpose of these sections is to present the analytic form
of Eq. (\ref{GeneralEstimator}).  We derive this expression by
computing the tensor ${\bf Q}$ and the Fisher matrix ${\bf F}$ using
Eq. (\ref{eqn:A}) combined with the results from
Sec. (\ref{sec:parm}).  We re-visit this derivation, which was
initially presented in \citetalias{2013MNRAS.433.2857M}, because we
find a number of corrections to the final form of the quadratic
estimator: we show that the most general expression for $\widehat{N}$
presented in \citetalias{2013MNRAS.433.2857M} is biased, such that,
$\langle \widehat{N}_i \rangle \ne N^{\rm true}_i$. However, we note
that in the limit where shot-noise dominates (labelled the `Schur
limit' by \citetalias{2013MNRAS.433.2857M}) their expression for the
estimator becomes unbiased.  We present the updated result for the
harmonic-space estimator in Eq. (\ref{estimatorfull}) and its
extension to configuration-space in Eq. (\ref{eqn:fullconfig}).  We
note that none of the numerical calculations in
\citetalias{2013MNRAS.433.2857M} are affected, since these require the
Fisher matrix and not the form of the estimator.

Our revisions can be understood as follows. First, our general form of
the quadratic estimator (i.e., Eq. (\ref{GeneralEstimator})) differs
from that presented in \citetalias{2013MNRAS.433.2857M}:
\begin{equation}
 \underbrace{ 
\widehat{N}_i = [\widehat{N}_i]_{\rm last} + \frac{1}{2}
\sum_j[{\bf F}^{-1}]_{ij} \bigg[ {\bf x}^{T}{\bf Q}_j{\bf x} -
  {\rm Tr}( {\bf A}^{-1} {\bf A}_{,j}) \bigg] \, .}_\text{Eq. (16) from \citetalias{2013MNRAS.433.2857M}}
\label{Eqn:mc}
\end{equation}
Importantly, Eq. (\ref{Eqn:mc}) is a simplified version of
Eq. (\ref{GeneralEstimator}).  The two expressions agree because of
the relation ${\rm Tr}({\bf Q}_j {\bf A}) = {\rm Tr}({\bf A}^{-1}
  {\bf A}_{,j})$. In order to simplify the derivation,
  \citetalias{2013MNRAS.433.2857M} neglect all derivatives of $A_{00}$
  -- note, we also make this approximation.  And such terms
  ($A_{00,i}$) occur in the expression for ${\bf Q}$.  Thus, by
  dropping these terms, one is making an approximation of ${\bf Q}$,
  ${\bf Q}^{\rm approx}$.  This approximation breaks the relation that
  equates Eq. (\ref{Eqn:mc}) to Eq. (\ref{GeneralEstimator}). So,
  ${\rm Tr}({\bf Q}^{\rm approx}_j {\bf A}) \ne {\rm Tr}({\bf A}^{-1}
    {\bf A}_{,j})$.  Therefore, when neglecting the derivatives of
    $A_{00}$, Eq. (\ref{GeneralEstimator}) needs to be the starting
    point of the derivation. A number of non-trivial corrections to
    the estimator proposed by \citetalias{2013MNRAS.433.2857M} result
    from this starting point.

\subsection{Harmonic space estimator}

To simplify the expressions that follow, using the notation of
\citetalias{2013MNRAS.433.2857M} we define the `Schur' parameter $S$
as
\begin{equation}
  S \equiv A_{00} \left(A_{00} - \sum_i \frac{A_{0
      i}^2}{A_{ii}}\right)^{-1} = \left(1 - \sum_i r_i^2 \right)^{-1}
  \, ,
  \label{eqn:schur}
\end{equation}
where the coefficents of ${\bf A}$ are defined in
Eq. (\ref{eqn:A}). Additionally, we define $r_i(\ell)$ as the
cross-correlation coefficient between the photo-$z$ sample and the
$i^{\rm th}$ redshift bin of the spec-$z$ sample: $r_i(\ell) \equiv
A_{0i}/(A_{00} \, A_{ii})^{1/2}$. Adopting these definitions, the full
estimator can be written as
\begin{eqnarray}
\widehat{N}^{(p)}_i &=& [\widehat{N}^{(p)}_i]_{\rm last} + \sum_j
(F^{-1})_{ij} \sum_{\ell, m} \, \left( \frac{S}{A_{00} \, A_{jj}}
\right) \frac{\partial A_{0j}}{\partial p_j} \nonumber \\
&\bigg[& \sum_k \left( \delta^K_{jk} + 2 S r_j r_k
  \sqrt{\frac{A_{jj}}{A_{kk}}} \right) ( \widehat{p} \; \widehat{s}_k
  - A_{0k} ) \nonumber \\
&-& \sum_k \frac{A_{0k}}{A_{kk}} \left( \delta^K_{jk} + S r_j r_k
  \sqrt{\frac{A_{jj}}{A_{kk}}} \right) ( \widehat{s}_k \;
  \widehat{s}_k - A_{kk} ) \nonumber \\
&-& \frac{S A_{0j}}{A_{00}} ( \widehat{p} \; \widehat{p} -
  A_{00} ) \bigg] \, ,
\label{estimatorfull} 
\end{eqnarray}
where $\delta^K_{ij}$ is the Kronecker delta.  In this expression the
($\ell,m$) dependence of the multipole coefficients is implicit, so,
$\widehat{s}_k = \widehat{s}_k(\ell,m)$ and $\widehat{p} =
\widehat{p}(\ell,m)$.

One can check this expression converges to the input theory as
follows.  First, we write the correction term as $\delta N_i \equiv
\widehat{N}^{(p)}_i - [\widehat{N}^{(p)}_i]_{\rm last}$.  Now,
assuming the input theory is correct, $\langle \widehat{p} \;
\widehat{s}_i \rangle = A_{0i}$, $\langle \widehat{p} \; \widehat{p}
\rangle = A_{00}$, and $\langle \widehat{s}_i \; \widehat{s}_j \rangle
= A_{ij}$. Then, following some algebra, one can show
Eq. \ref{estimatorfull} implies $\langle \delta N_i \rangle = 0$,
thus, proving that the estimator will converge.  For the equivalent
equation from \citetalias{2013MNRAS.433.2857M}, one can show $\langle
\delta N_i \rangle \ne 0$.

The Fisher matrix (Eq. \ref{eqn:fish1}) remains unchanged from
\citetalias{2013MNRAS.433.2857M}, where
\begin{equation}
F_{ij} = \sum_{\ell, m} \frac{S}{A_{00}}
\left(\frac{\delta^K_{ij}}{A_{ii}} + 2\,S \, \sqrt{\frac{r_i^2
    r_j^2}{A_{ii} \, A_{jj}}} \right) \, [A_{0i}]_{,i} \;
     [A_{0j}]_{,j} \, .
\label{eqn:limber_fish}
\end{equation}
In the limit where shot noise dominates (i.e., where $r_i(\ell)\approx
0$ and $S\approx 1$) and neglecting auto-correlations, our result
(Eq. \ref{estimatorfull}) agrees with Eq. (36) from
\citetalias{2013MNRAS.433.2857M}.

\begin{figure*}
\centering
\includegraphics[width=15.0cm]{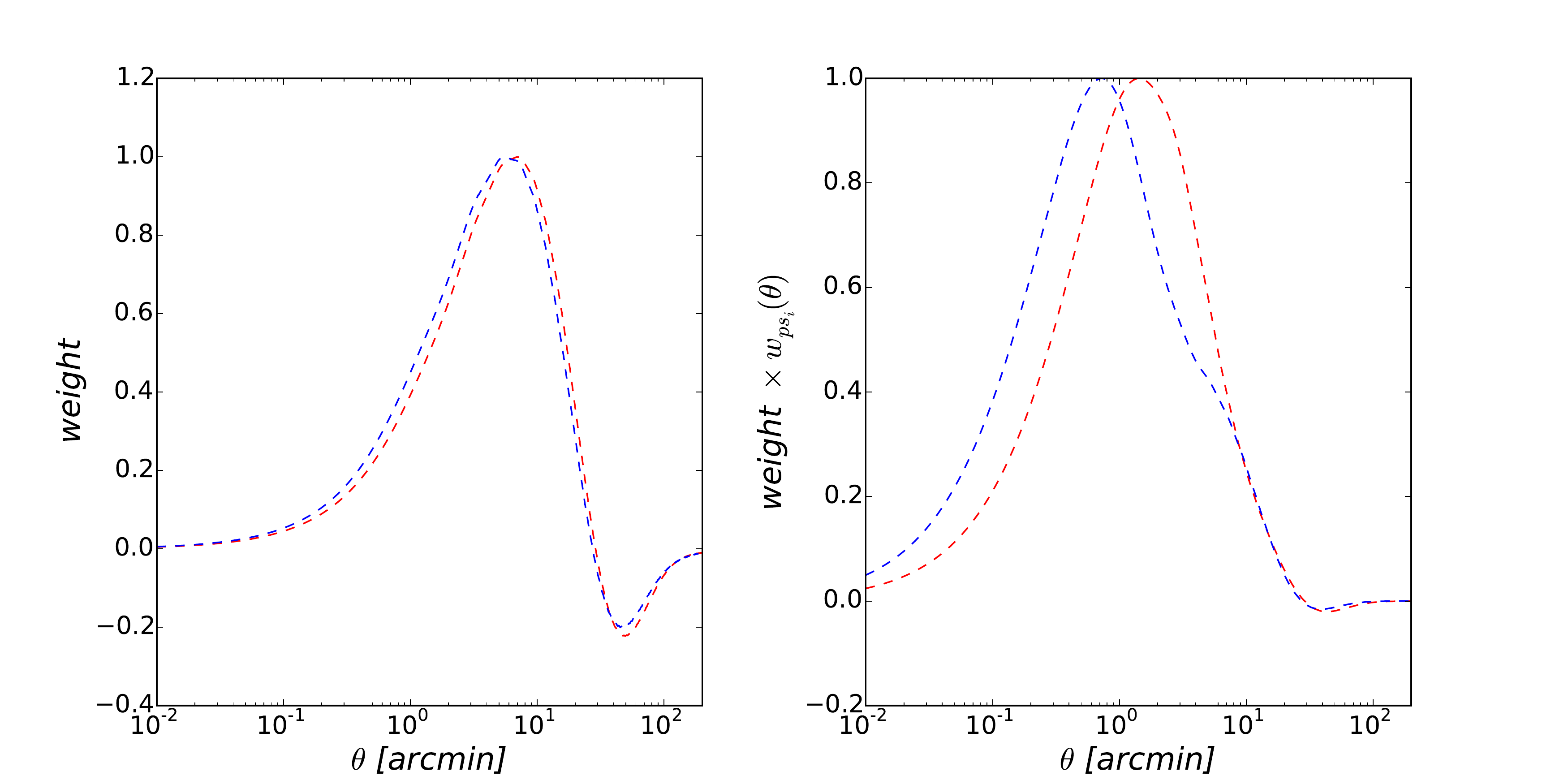}
\caption{The weights $D_i(\theta)$ defined in Sec. (\ref{sec:config})
  (left panel) and the combination $D_i(\theta) \times
  w_{ps_i}(\theta)$ (right panel) for the mock catalogue tests.  The
  weights and correlation functions are derived for the $z=0.225$
  (blue) and $z=0.525$ (red) redshift bins, for illustration.  The
  combination $D_i(\theta) \times w_{ps_i}(\theta)$ represents the
  angular scales used by the estimator to measure the redshift
  probability distribution, in accordance with
  Eq. (\ref{eqn:fullconfig}).}
\label{weights}
\end{figure*}

\subsection{Configuration space estimator}
\label{sec:config}

When analyzing observational data we will work exclusively in
configuration space, which allows us to avoid difficulties with
complex survey geometries (in the future this may not be necessary
\citep[see,][]{2016MNRAS.455.4452A,2016MNRAS.456.1508K}).  Thus, in
order to match our theory with observations, we need to convert our
estimator in Eq. (\ref{estimatorfull}) from harmonic to configuration
space.  This conversion is simplified by the following relation
\citepalias{2013MNRAS.433.2857M}:
\begin{align}
 \sum_{l,m}v_i(l) \bigg [ \widehat{p}(l,m)&\widehat{s}_i(l,m)- N^{(p)}_i \bigg ] \nonumber \\
& = 8 \pi^2 \int dx \, v_i(x) \, \widehat{w}_{p s_i}(x) \nonumber \\
& \approx 8 \pi^2 \sum_{\alpha} \Delta \theta_{\alpha} \theta_{\alpha} \, v_i(\theta_{\alpha}) \, \widehat{w}_{p s_i}(\theta_{\alpha})  \, ,
\label{eqn:convertTheta}
\end{align}
where $\widehat{w}_{p s_i}(\theta)$ is the observed angular
cross-correlation function, $\theta$ is the angular separation scale,
and $x = \hat{n} \cdot \hat{n}^{\prime} \equiv \cos \theta$.  Here we
have explicitly subtracted the shot-noise component\footnote{We note
  that incorrectly modelling the shot-noise component will introduce a
  bias into the final measurements.}.  Eq. (\ref{eqn:convertTheta}) is
valid for an arbitrary function ${\bf D}{(l)}$, which is related to
${\bf D}(\theta)$ by
\begin{equation}
D_i(\theta) = \sum_\ell \left( \frac{2\ell+1}{4\pi} \right) \,
D_i(\ell)\, P_{\ell}(\cos{\theta}) \, ,
\label{eqn:convertTheta2}
\end{equation}
where $P_\ell$ are the Legendre polynomials.

Our measurements of the angular correlation functions are made in bins
of width $\Delta \theta_{\alpha}$, with central values
$\theta_{\alpha}$. These values set the properties of the summation in
Eq. (\ref{eqn:convertTheta}). Note, because the kernel $(=
\theta_{\alpha} \, v_i(\theta_{\alpha}) \, \widehat{w}_{p
  s_i}(\theta_{\alpha}))$ is not a slowly varying function, a narrow
$\theta$ spacing $(\Delta \theta_{\alpha})$ is needed to accurately
approximate this integral.

Now, to convert Eq. (\ref{estimatorfull}) into configuration space
we first re-write the estimator in terms of four weighting functions
defined as
\begin{align}
D_i(l) &\equiv \left( \frac{S}{A_{00} \, A_{ii}} \right) \,
\frac{\partial A_{0i}}{\partial p_i} \,\, , \\ E_{ij}(l) &\equiv
D_i(l)\times 2 S r_i r_j\sqrt{\frac{A_{ii}}{A_{jj}}} \, ,
\end{align}
and
\begin{equation}
H_i(l) \equiv D_i(l)\times \frac{A_{0i}}{A_{ii}} \,\, , G_i(l)
\equiv D_i(l) \times S \frac{A_{0i}}{A_{00}} \, .
\end{equation}
Finally, using Eq. (\ref{eqn:convertTheta}) one finds that our estimator
in configuration space takes the form
\begin{align}
\widehat{N}^{(p)}_i& = [\widehat{N}^{(p)}_i]_{\rm last} + 8 \pi^2 \sum_j (F^{-1})_{ij}  \label{eqn:fullconfig} \\
& \sum_{\alpha} \Delta \theta_{\alpha} \, \theta_{\alpha}\Bigg [ \nonumber
D_j(\theta_{\alpha}) \,\bigg\{ \widehat{w}_{ps_j}(\theta_{\alpha}) - w_{ps_j}(\theta_{\alpha}) \bigg\} \nonumber \\
& + H_j(\theta_{\alpha}) \, \widehat{w}_{s_js_j}(\theta_{\alpha})-  G_j(\theta_{\alpha}) \, \widehat{w}_{pp}(\theta_{\alpha})  \nonumber \\
&+ \sum_{k} E_{jk}(\theta_{\alpha})\bigg\{ \widehat{w}_{p s_{k}}(\theta_{\alpha}) - w_{ps_k}(\theta_{\alpha}) + \frac{1}{2}\widehat{w}_{s_ks_k}(\theta_{\alpha}) \bigg\} \Bigg ] \nonumber \, ,
\end{align}
where $\{{\bf D,E,G,H}\}$ have been transformed to configuration space
using Eq. (\ref{eqn:convertTheta2}).  Ignoring both the bin-to-bin
correlations, such that the Fisher Matrix is diagonal, and the
auto-correlation terms, the second term in Eq. (\ref{eqn:fullconfig})
becomes \citepalias{2013MNRAS.433.2857M}
\begin{align}
8 \pi^2 (F^{-1})_{ii}\sum_{\alpha} \Delta \theta_{\alpha} \, \theta_{\alpha} \Bigg [ \nonumber D_i(\theta_{\alpha}) \,\bigg\{ \widehat{w}_{ps_i}(\theta_{\alpha}) - w_{ps_i}(\theta_{\alpha}) \bigg\} \Bigg ] \, .
\end{align}
The result is now much more intuitive. The $N(z)$ is reconstructed
from a weighted minimization of $\{ {\widehat
  w}_{ps_i}(\theta_{\alpha})-w_{ps_i}(\theta_{\alpha})\}$, where the
weights are given by $D_i(\theta)$.  Note that the scales that
contribute most to the estimator are represented by the combination
$D_i(\theta)w_{ps_i}(\theta)$.

To illustrate the angular sensitivity of the estimator, in
Fig. (\ref{weights}) we plot these weights for the mocks introduced in
the next section.  We find that the weights peak at $\theta \sim 2$
arcmins.  However, as emphasized by \citetalias{2013MNRAS.433.2857M},
the breadth and smoothness of the weighting function implies that the
Fourier modes being traced are on quasi-linear scales: sharp cuts in
angle have a greater sensitivity to non-linearity than a smoother
filter.

\section{Application to simulations}
\label{sec:simtests}

In this section, we test our methodology using 20 sets of mock halo
catalogues\footnote{We find $20$ mock catalogues is sufficient for the
  level of error we wish to investigate.} created from the N-body
simulations described in Sec. (\ref{sec:simulations}).  For each 60
deg$^2$ simulation we generated a uniform redshift distribution of
mock spectroscopic sources within the range $0.1 < z < 0.9$, adopting
an angular density of 1000 sources deg$^{-2}$.  In addition, we
sampled mock photometric galaxies using a Gaussian redshift
distribution with mean 0.5 and standard deviation 0.1 with a density
of 1 source arcmin$^{-2}$ (which roughly mimicks a typical tomographic
bin in KiDS). For the purposes of this test we generated each sample
by randomly sampling from the halo catalogue at each redshift, such
that the bias factors of the photometric and spectroscopic samples are
expected to be the same (this would not necessarily be true for a real
data sample).  We then performed a cross-correlation analysis dividing
the spectroscopic sources into 16 redshift bins of width $\Delta z =
0.05$.  For further details on the mock catalogues we refer the reader
to \citet{2016MNRAS.462.4240B}, section 6.1.

We note that each individual mock provides constraints comparable in
precision to the observational datasets used in our analysis (each
mock realization contains $\sim 60{,}000$ spec-$z$ galaxies, compared
to $\sim 40{,}000$ spec-$z$ galaxies in 2dFLenS).

\subsection{Auto- and cross-correlation measurements}

We measured the angular auto- and cross-correlation functions in our
analysis using the Landy-Szalay estimator \citep{1993ApJ...412...64L},
generating random catalogues 10 times larger than the data
distribution.  For example, the cross-correlation between samples $i$
and $j$ is
\begin{equation}
\label{eq:LS}
  \begin{split}
    w_{i,j}(\theta) = \frac{(D_iD_j)_{\theta}}{(R_iR_j)_{\theta}}
    \frac{N_{R,i}N_{R,j}}{N_{D,i}N_{D,j}} -
    \frac{(D_iR_j)_{\theta}}{(R_iR_j)_{\theta}}
    \frac{N_{R,i}}{N_{D,i}} \\
    - \frac{(D_jR_i)_{\theta}}{(R_iR_j)_{\theta}}
    \frac{N_{R,j}}{N_{D,j}} + 1 \, ,
  \end{split}
\end{equation}
where $(D_iD_j)_{\theta}$, $(D_iR_j)_{\theta}$ and $(R_iR_j)_{\theta}$
are the respective pair counts between the two data samples, the data
and random samples, and the two random samples, as a function of the
angular separation $\theta$.  The number counts of the data sample $i$
and the random sample $j$ are $N_{D,i}$ and $N_{R,j}$, respectively.
The correlation functions are all measured using 30 equally
logarithmically-spaced angular bins between 0.01 and 1 deg.  We
estimate the errors in the measurements using jack-knife re-sampling,
although these errors are not used in the quadratic estimation
process.

For each mock catalogue we measured the following statistics: The
auto-correlation of the photometric galaxies, $w_{pp}(\theta)$; the
$16$ spectroscopic auto-correlations in each redshift bin,
$w_{s_is_i}(\theta)$; and the $16$ photometric-spectroscopic
cross-correlations, $w_{ps_i}(\theta)$, where the index $i$ runs
across the $16$ redshift bins of the spectroscopic sample.

\subsection{Applying the estimator to the mocks}
\label{sec:resultscrosscorr}

In this section we apply the quadratic estimator to the mock
catalogues and infer results for the redshift probability distribution
$P^{(p)}(z)$ across the spectroscopic bins.  For the purposes of this
test we only use the angular cross-correlation functions as inputs to
the quadratic estimator.  Thus, we apply Eq. (\ref{eqn:fullconfig})
and drop the auto-correlation terms.  We note that in this limit the
estimator remains unbiased, as the auto-correlation terms cancel.  As
discussed in the next section, we use the auto-correlations of the
spec-$z$ sample separately, to measure the redshift evolution of the
galaxy bias.

Every iteration of the quadratic estimator returns a correction term
to the probability distribution $\delta P_i$ computed from the
previous best-guess estimate of the redshift distribution.  We define
the estimator to be {\it converged} once the condition $\sum_i \delta
P_i < 5 \times 10^{-3}$ is met.  At this level of accuracy, the
estimated correction is on average an order of magnitude smaller than
the error for a given redshift bin, i.e., $\delta P_i/\sigma(P_i)
\approx 0.1$.

After each iteration of the quadratic estimator, we also enforce the
normalization condition $\sum_i P_i = 1$.  This produces an overall
amplitude shift that is minimal in most cases, although we do expect a
small bias to be introduced when imposing this constraint because the
normalization condition holds only for the underlying $P_i$, not the
estimated $\hat{P}$.

\subsection{Convergence of the estimator}

\begin{figure*}
\centering
\includegraphics[width=15.0cm]{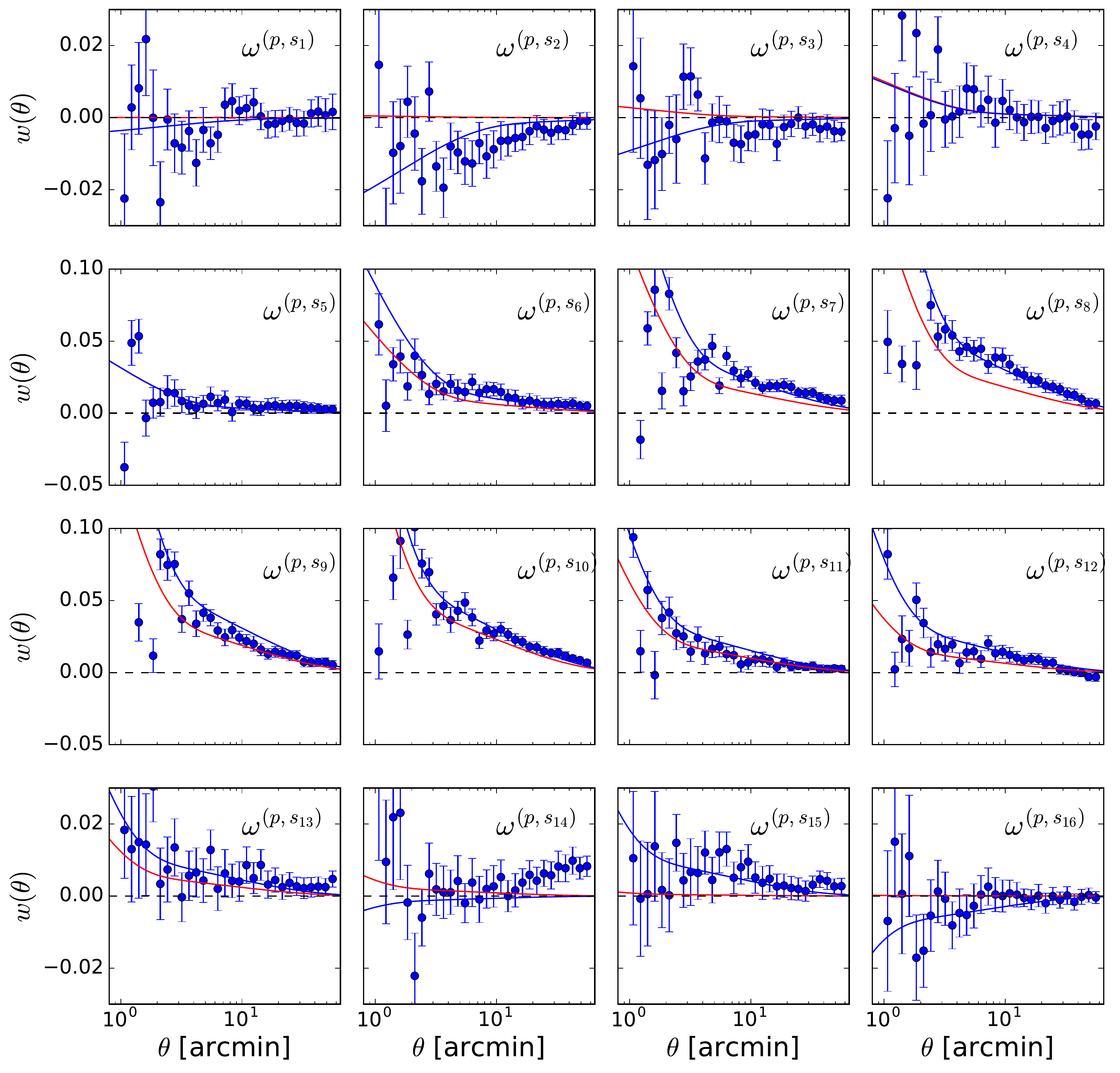}
\caption{Examination of the convergence of the estimator.  The blue
  points show the measured angular cross-correlation functions for a
  single mock catalogue for all the spectroscopic redshift bins.  Each
  panel corresponds to a separate redshift bin and $\omega^{(p,s_{\rm
      i})}$ indicates a cross-correlation between the photo-$z$ sample
  and the $i^{\rm th}$ bin in the spec-$z$ sample. The red lines show
  the predictions of the cross-correlation functions using the true
  underlying $P(z)$ -- a Gaussian with mean 0.5 and standard deviation
  0.1.  The blue lines show our predictions using the inferred ${\hat
    P}(z)$ from the quadratic estimator, which are seen to closely
  track the measurements.  We note the different $y$-axis ranges in
  the panels.}
\label{crosscorr_test}
\end{figure*}

\begin{figure*}
\centering
\includegraphics[width=15.0cm]{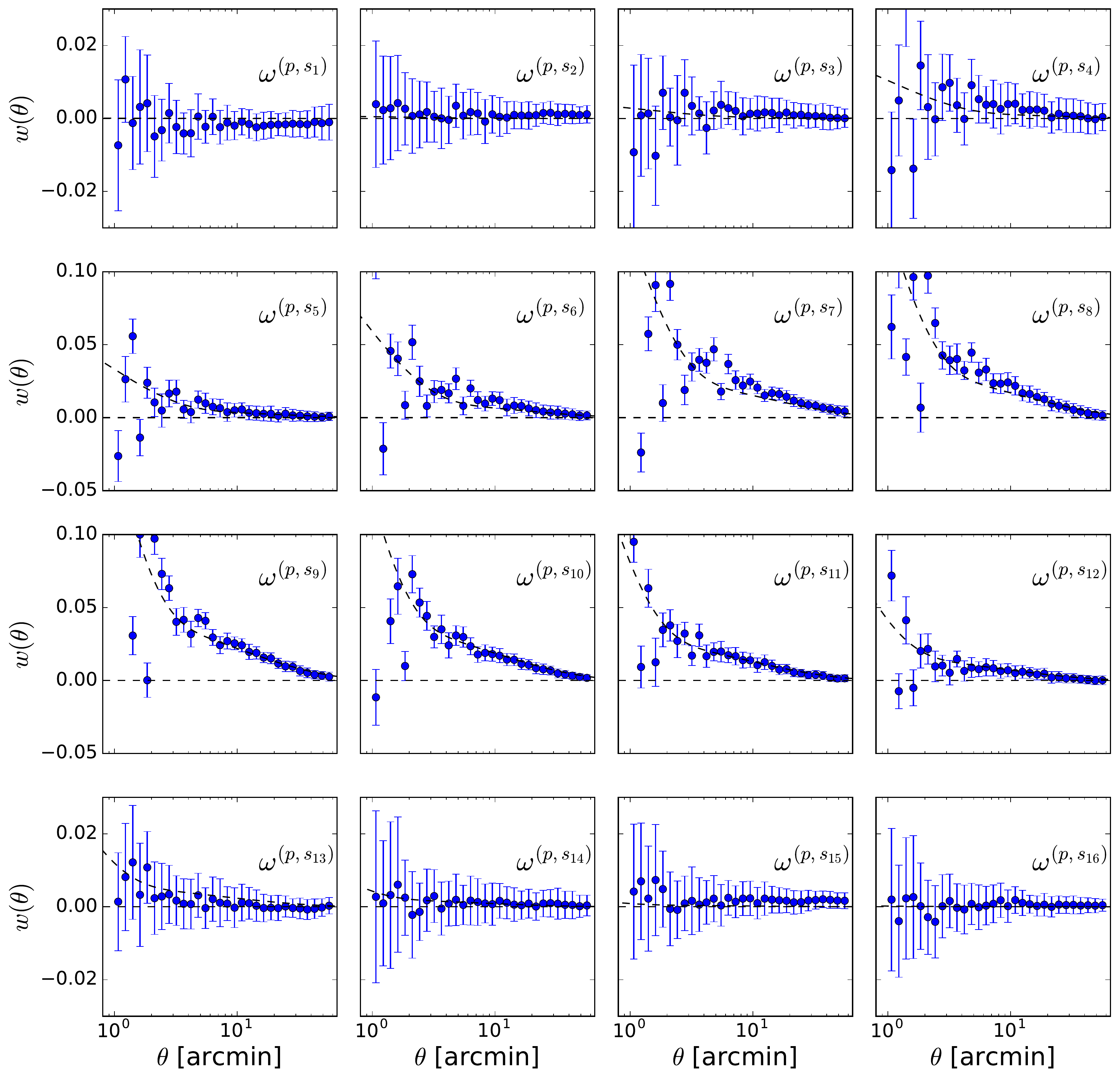}
\caption{Test of the accuracy of our cross-correlation model.  The
  blue points show the mean angular cross-correlation function
  measured from 20 mock catalogues for all the spectroscopic redshift
  bins, with the error taken from a single mock.  The red lines show
  our model predictions, using the cosmological parameters of the
  simulation and the mean bias factors.  Each panel corresponds to a
  separate redshift bin and $\omega^{(p,s_{\rm i})}$ indicates a
  cross-correlation between the photo-$z$ sample and the $i^{\rm th}$
  bin in the spec-$z$ sample. The bar, $\bar{\omega}$, is simply a
  reminder that we are averaging over mocks when determining these
  results.}
\label{model_test}
\end{figure*}

We can assess the convergence of the estimator and the accuracy of the
modelling by comparing the observed, reconstructed and model angular
cross-correlation functions obtained from a single (representative)
mock catalogue, as shown in Fig. (\ref{crosscorr_test}).  Each panel
within this figure illustrates the cross-correlation function
$w_{ps_i}(\theta)$ for one of the spectroscopic redshift bins,
starting from the lowest redshift in the top left-hand corner.  We
overplot the model prediction for the angular cross-correlations as
the red lines, derived using Eq. (\ref{wtheta2}) and using the bias of
the samples determined as discussed in Sec. (\ref{subsec:bias}) below.
The blue lines in Fig. (\ref{crosscorr_test}) show the {\it
  reconstructed} angular cross-correlation functions, which we obtain
by using the recovered photo-$z$ redshift distribution,
$\hat{P}^{(p)}_i$, to compute the angular cross-correlation via
\begin{equation}
{\widehat w}_{ps_i}(\theta) = b^{(s)}_i b^{(p)}_i P^{(s)}_i
\hat{P}^{(p)}_i w_{s_i s_i}(\theta) \, .
\end{equation}
We can now assess the convergence of the estimator by comparing the
reconstructed predictions (blue lines) to the mock measurements (blue
points).  Thus, from Fig. (\ref{crosscorr_test}), we observe that the
combination $\{ {\widehat
  w}_{ps_i}(\theta_{\alpha})-w_{ps_i}(\theta_{\alpha})\}$ is being
successfully minimized.  One should keep in mind the effective
$\theta$-dependent weights being applied to this minimization, as
shown by Fig. (\ref{weights}).  Comparing cross-correlation functions
is a useful validation of the estimator, as this test is less
sensitive to inaccuracies in modelling the correlation functions and
galaxy bias than comparing the inferred $P^{(p)}_i$ distribution to
the input\footnote{For example, in a situation where the model
  over-predicts the amplitude of the mock correlation function, the
  estimator will respond by reducing $P^{(p)}_i$ relative to the true
  value.  However, in this case ${\widehat w}_{ps_i}$ should still
  agree with the simulation prediction.}.

\subsection{Accuracy of the cross-correlation model}
\label{subsec:acc}

Systematic errors in the modelling of the cross-correlation statistics
-- for example, due to the breakdown of an assumption such as linear
galaxy bias -- could propagate into a bias in the inferred photometric
redshift distribution.  We assessed this potential source of
systematic error by comparing the average of the mock
cross-correlation function measurements to our model predictions, the
results of which are shown in Fig. (\ref{model_test}).  The models are
generated using the galaxy bias determined for the spectroscopic
sample in Sec. (\ref{subsec:bias}) below, and in this figure we plot
the average error for individual mocks.  We conclude that our
modelling of the angular cross-correlation function is sufficient at
the level of statistical errors present in a single mock catalogue.

However, if we instead use the error in the mock mean (dividing by
$\sqrt{20}$) we observe some tension between the simulation
predictions and the analytic modelling of the correlation functions.
These tests reveal that there are non-linear effects in the mocks not
captured by our model (e.g., non-linear galaxy bias), and also that
the jack-knife errors do not fully capture the scatter in the
measurements.  These issues could be mitigated by restricting our
analyses to larger scales, and by deducing the statistical errors
using a dispersion in the mocks, rather than by jack-knife techniques.
However, we leave such investigations to future work.

\subsection{Measuring the galaxy bias factors}
\label{subsec:bias}

\begin{figure*}
\centering
\includegraphics[width=15.0cm]{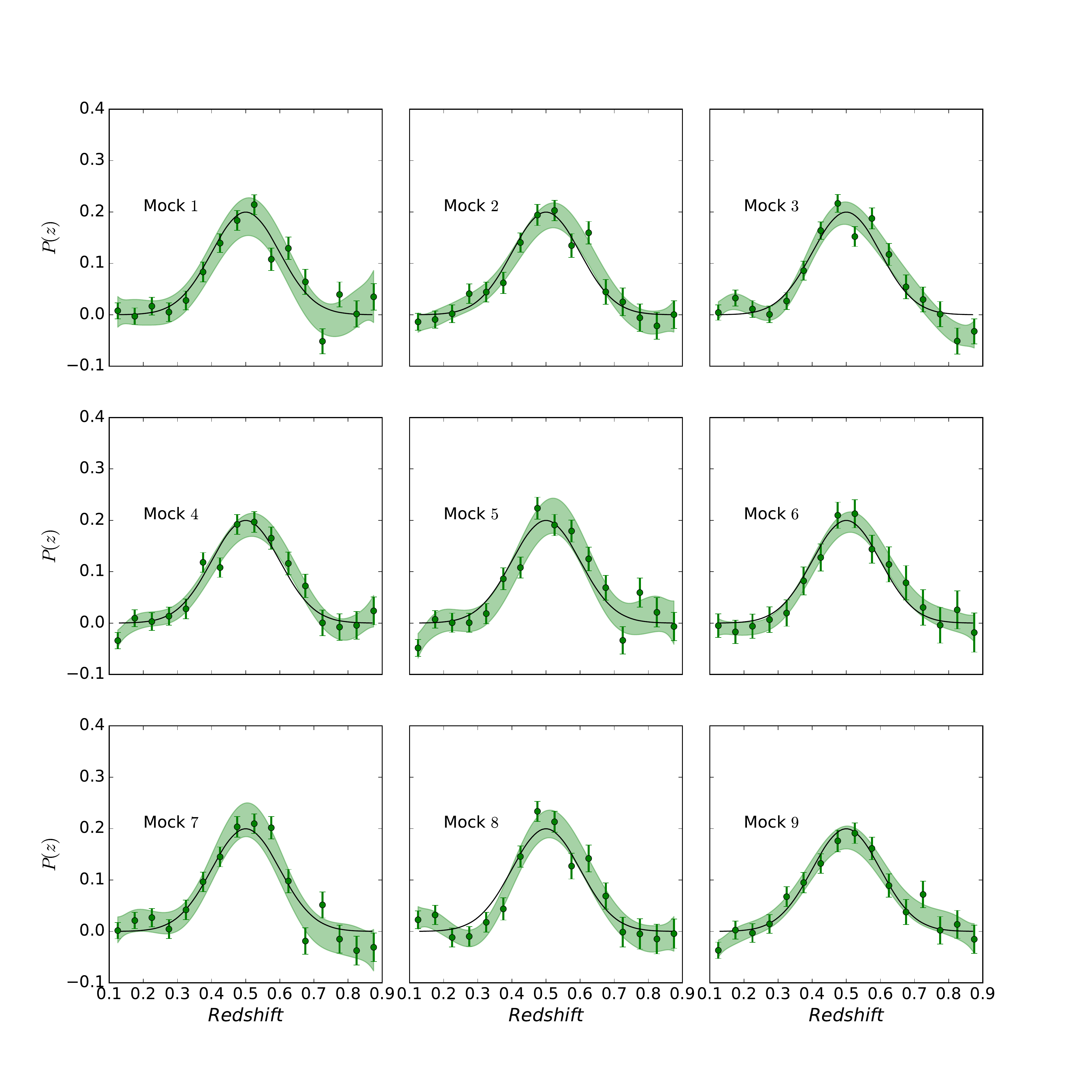}
\caption{The inferred redshift probability distribution for nine
  independent mock catalogues.  The green points show the
  reconstructed galaxy redshift distributions estimated by quadratic
  estimation (${\hat P}$), with error bars derived from the relevant
  Fisher matrix.  The green bands show the $95\%$ confidence intervals
  for a Gaussian Process model trained using the green points.  The
  black line shows the input redshift probability distribution, which
  is recovered with reasonable accuracy.}
\label{mock_results}
\end{figure*}

In order to test whether our quadratic estimation pipeline recovers
the input source redshift distribution of the mocks, we also require
the redshift evolution of the galaxy bias factors, which we have
arranged (by sampling halos in the same mass range) to be the same for
the spectroscopic and photometric samples.  We determined the redshift
evolution of this bias using the auto-correlation function
measurements of each spec-$z$ sample, using chi-squared minimization
to fit for $b^{(s)}_i$.

We note that the effects of noise will cause fluctuations between
$b^{(s)}_i$ and $b^{(p)}_i$.  In particular, given the significantly
lower number density of spec-$z$ galaxies relative to the photo-$z$
sample, the bias measurements of the spec-$z$ sample will be less
accurate.  We overcome this by averaging the $b^{(s)}_i$ values over
the 20 mocks.

We propagate the noise in the measurement of $b^{(s)}_i$ into the
inferred redshift probability distributions by empirically determining
that the scatter in the bias measurements across mocks, $\sigma(b)
\sim 0.1$, produces a scatter in the probability distribution
$\sigma(P) \sim 0.002$.

\subsection{Reconstructed mock redshift distributions}

\begin{figure*}
\centering
\includegraphics[width=15.0cm]{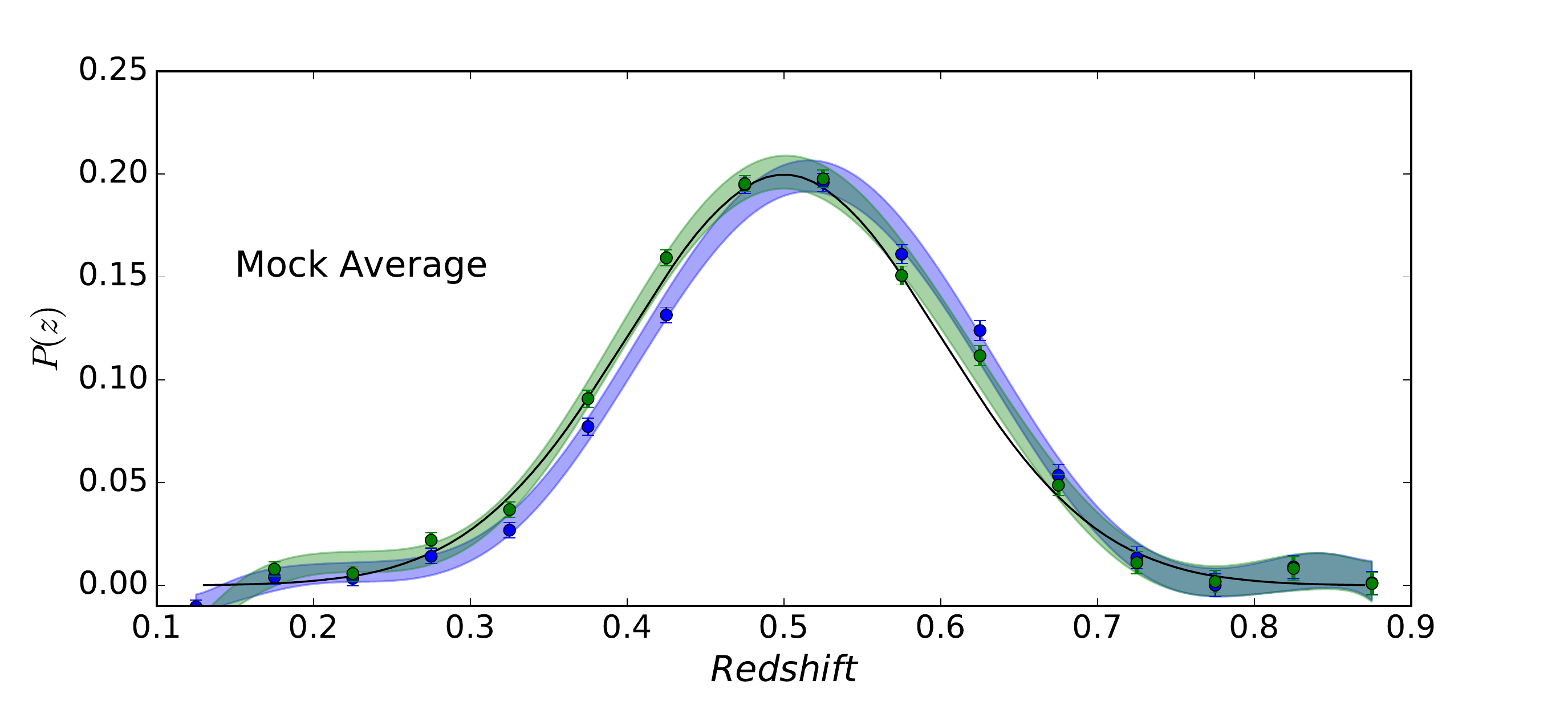}
\caption{The average of the reconstructed redshift probability
  distributions over 20 mock catalogues.  The blue points show the
  estimated redshift distribution ${\hat P}$ when we set the bias of
  the photo-$z$ sample to the mean value of the spec-$z$ bias
  values. The green points show the estimated ${\hat P}$ when we set
  $b^{(p)}(z) = 1.0 + \alpha(z - z_0)$, where $\alpha = 2.2$ and $z_0
  = 0.45$. The black line shows the input redshift probability
  distribution, which is recovered more accurately by the second
  method.}
\label{mock_results_ave}
\end{figure*}

\begin{figure*}
\centering
\includegraphics[width=15.0cm]{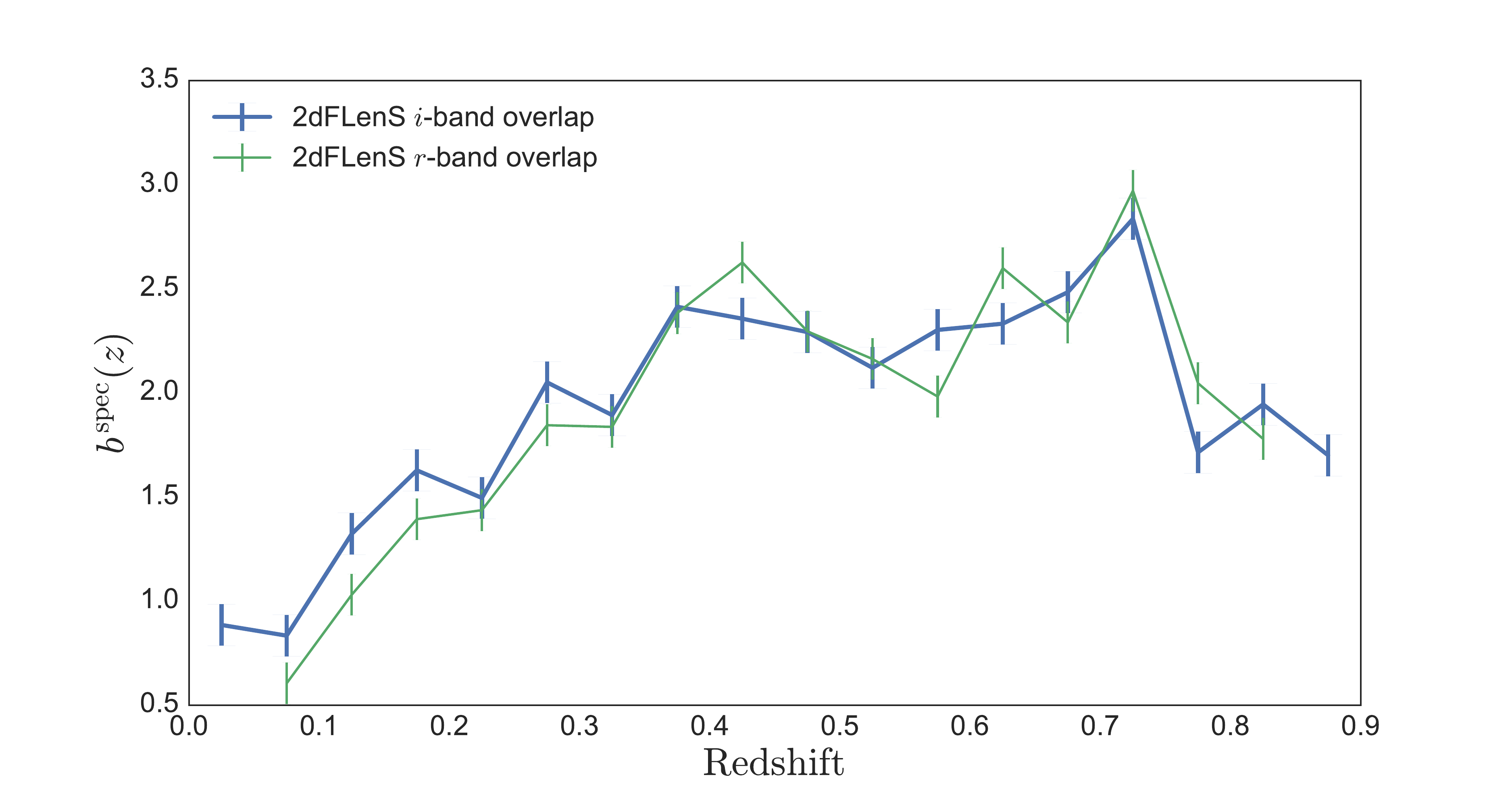}
\caption{The bias evolution of the spectroscopic 2dFLenS sample for
  the overlaps with the KiDS $r$-band and $i$-band samples, measured
  from the galaxy auto-correlations.  The errors are determined by
  scaling the results from mock catalogues.}
\label{bias_evolv}
\end{figure*}

We now consider the results for the recovered redshift distributions
of the mock source catalogues, measured in 16 step-wise redshift bins
of width $\Delta z = 0.05$.  We present a random subset of these
results for 9 mocks in Fig. (\ref{mock_results}).  The green data
points display the best-fitting $P^{(s)}_i$ values and the 1-$\sigma$
errors are derived from the Fisher matrix.  The black line shows the
Gaussian redshift distribution assumed in the simulations.  We can see
that the reconstruction has proved generally successful.

For applications to weak gravitational lensing, a flexible functional
form for the redshift distribution is more convenient than a step-wise
binning.  We adopt {\it Gaussian processes} (GPs) as a method to infer
such functional forms for redshift distributions.  Briefly, GPs
provide {\it non-parametric} Bayesian modelling for supervised
learning problems.  For details, we refer the reader to
\citet{2013arXiv1311.6678S}.  GPs also readily allow for the inclusion
of a prior on the smoothness of the reconstructed function, naturally
expected from a survey selection function.\footnote{We are not
  implying here that the final $P(z)$ will necessarily be a very
  smooth function of redshift, as photo-$z$ cuts can cause sharp
  variations.  Rather, we are implying that the amplitude between
  redshift bins will be highly correlated.  Thus, this setup still
  allows sharp fluctuations, although they are less probable.}

We build GP models using the python module {\tt SCIKIT-learn}
\citep{scikit-learn}.  As input one needs to define the functional
form for the adopted (redshift) correlation function $C(z,z^\prime)$
and set the {\it characteristic scale} $L$, which intuitively
determines the typical distance between peaks (i.e., the smoothness
scale) of the function.  We set the redshift correlation as a Gaussian
function
\begin{equation}
C(z,z^\prime) = \sigma_f^2
\exp\left[-\frac{(z-z^\prime)^2}{2L^2}\right] \, ,
\end{equation}
where $\sigma_f^2$ defines the variance of the function, and we set $L
= 1$.  The optimal choice of $L$ and $C$ will depend on the survey in
question.

In Fig. (\ref{mock_results}) we display the 2-$\sigma$ confidence
intervals for the GP models as the green band.  We observe that the
reconstructed GP distributions encompass the input mock redshift
distribution for the vast majority of mocks and redshift bins.  More
quantitatively, using all 20 mock catalogues we measure an average
$\chi^2$ of $19.95$ for 18 degrees of freedom (ignoring bin-to-bin
correlations) demonstrating that, at the level of statistical error of
{\it individual} mock catalogues (and hence of the observational
datasets used in our study), our measurements are statistically
consistent with the input distribution.

We now consider a more accurate test of our methodology, using the
average $P^{(p)}(z)$ values over all 20 mock catalogues.  For this
test, we consider two methods to estimate the bias of the photo-$z$
sample, the results of which are shown in
Fig. (\ref{mock_results_ave}).  For {\it method 1} we adopt our
default model, i.e., $b^{(p)}_i = \langle b^{(s)}_i \rangle$, where we
average the spec-$z$ bias factors over 20 mocks (blue points and
band).  For {\it method 2} we use a bias evolution model with two free
parameters defined as $b^{(p)}(z) = 1.0 + \alpha(z - z_0)$, where we
have fitted $\alpha = 2.2$ and $z_0 = 0.45$ from the mocks (green
points and band).  The input Gaussian redshift distribution is
displayed as the black line.

Inspecting Fig. (\ref{mock_results_ave}), we find a significant
discrepancy between our predictions based on method 1 and the input
distribution.  Qualitatively, we observe that for $z<0.5$ the
distribution tends to be underestimated, while for $z>0.5$ the
distribution tends to be overestimated.  We interpret these
discrepancies as a limitation of our modelling of the
cross-correlation function and the influence of non-linear galaxy
bias.  We find that the predictions using the second bias model are
more representative.

\section{Application to KiDS and 2dFLenS}
\label{sec:results}

\begin{figure*}
\centering
\includegraphics[width=15.0cm]{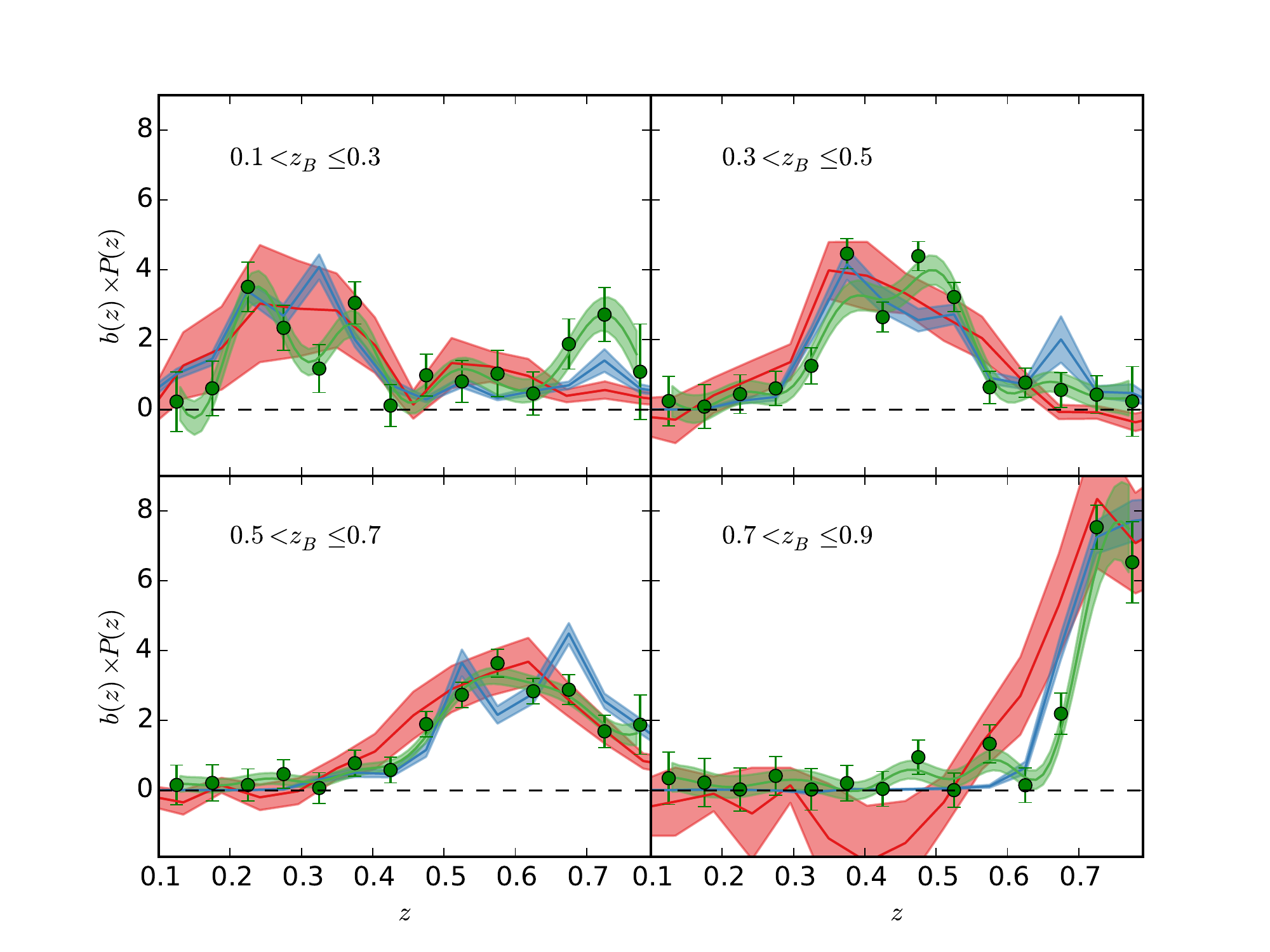}
\caption{Reconstruction of the combination $b^{(p)} P^{(p)}(z)$ by
  applying quadratic estimation to cross-correlations between the
  KiDS-450 $r$-band catalogue and 2dFLenS, for four tomographic bins
  of the photometric catalogue ($0.1 < z_B \le 0.3$, $0.3 < z_B \le
  0.5$, $0.5 < z_B \le 0.7$ and $0.7 < z_B \le 0.9$).  The green data
  points display the results of quadratic estimation, and the green
  bands show the $68\%$ confidence intervals for a Gaussian Process
  model trained using these measurements.  These results are compared
  with determinations by methods using small-scale cross-correlation
  (red bands) and weighted direct calibration (blue bands) (see
  \citet{2016arXiv160605338H} for more details about these methods).
  For the purposes of comparison, all the distributions have been
  normalized such that $\int_{0.1}^{0.8} dz \, b^{(p)} \, P^{(p)} =
  1$.}
\label{rband_results}
\end{figure*}

\begin{figure*}
\centering
\includegraphics[width=15.0cm]{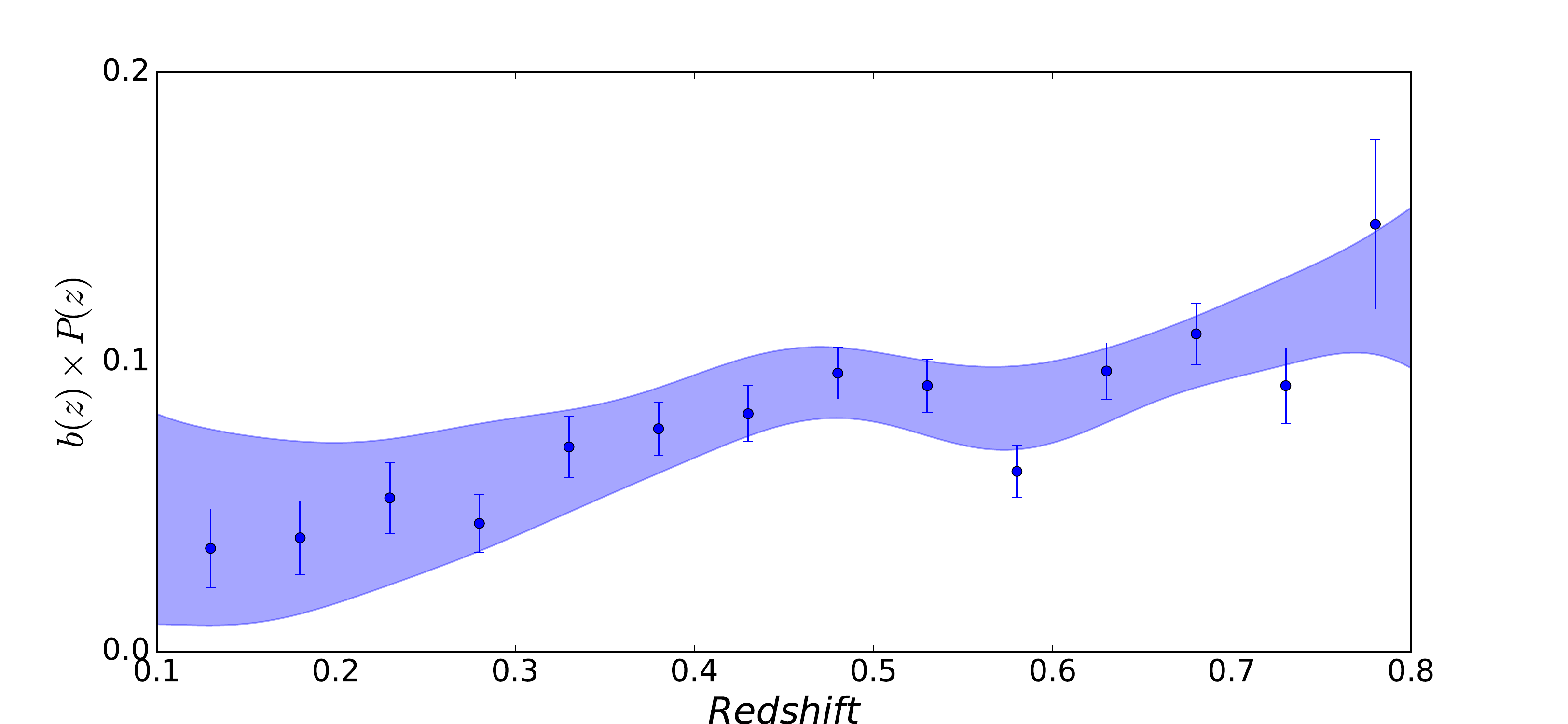}
\caption{Reconstruction of the combination $b^{(p)} P^{(p)}(z)$ by
  applying quadratic estimation to cross-correlations between the
  KiDS-800 $i$-band catalogue and 2dFLenS.  The blue points display
  the results of quadratic estimation, and the blue band shows the
  $95\%$ confidence intervals for a Gaussian Process model trained
  using these measurements.}
\label{iband_results}
\end{figure*}

In this section, we apply the quadratic estimation methodology to
infer the product $b^{(p)}(z) \, P^{(p)}(z)$ for sources detected in
the separate $r$-band and $i$-band catalogues of the KiDS imaging
survey, using their cross-correlation with the 2dFLenS spec-$z$
catalogue.  Because (unlike for the mock catalogues) we have no
information on $b^{(p)}(z)$, we cannot break the degeneracy
$b^{(p)}(z) \, P^{(p)}(z)$ without further assumptions.  We
cross-correlate the photometric samples with the 2dFLenS catalogue in
18 redshift bins in the range $0 < z < 0.9$.  The $r$-band and
$i$-band samples have a different degree of overlap with 2dFLenS, and
we use a total of $13{,}740$ and $25{,}443$ spec-$z$ galaxies for the
respective cross-correlations.  Following the cosmic shear analysis of
\citet{2016arXiv160605338H}, we divided the KiDS-450 $r$-band imaging
dataset into four tomographic bins based on the {\tt BPZ} redshift
$z_B$: $0.1 < z_B \le 0.3$, $0.3 < z_B \le 0.5$, $0.5 < z_B \le 0.7$
and $0.7 < z_B \le 0.9$.  Photo-$z$ information is not available for
the KiDS-800 $i$-band dataset, and we do not sub-divide it.  In the
correlation function measurement, we now weight each source by its
optimal weight in the estimation of shear statistics, such that we are
constraining the weighted redshift probability distribution of the
sources \citep{2013MNRAS.429.2858M}.

Since we are only utilizing spectroscopic data in the range $z < 0.9$,
we cannot derive the full KiDS source redshift distributions for most
of the samples.  Rather, our motivation is to demonstrate an
application of our methodology on a realistic dataset.

We fit for the spec-$z$ bias values in each redshift bin as outlined
in the previous section, and show the results in
Fig. (\ref{bias_evolv}).  We perform a rough scaling of the expected
error in the bias compared to the mocks, and propagate this error into
the determination of $b^{(p)} \, P^{(p)}$ by adding a term $\sigma(P)
= 0.02$ to the standard Fisher matrix errors.

It is possible for measurements of the angular cross-correlation
function to be negative due to either noise, or effects such as
incompleteness in the imaging catalogue around bright spectroscopic
sources or other systematics.  These points are unphysical in our
model, which predicts the cross-correlation functions by scaling the
auto-correlations, which are positive definite.  To address this
problem we add a positive definite prior which effectively shifts
negative $b^{(p)} \, P^{(p)}$ values to zero.

In Fig. (\ref{rband_results}) we show the reconstructed $b^{(p)}(z)
\times P^{(p)}(z)$ measurements for each tomographic bin of the
KiDS-450 $r$-band data, in comparison with other determinations of
this distribution presented by \citet{2016arXiv160605338H}.  The green
data points show the quadratic estimation with 1-$\sigma$ error bars,
and the shaded green band gives the $68\%$ confidence interval for the
GP model.  Only 14 redshift bins are shown per tomographic bin; it was
not possible to recover an estimate for the first two and last two
redshift bins because of the low number of spec-$z$ galaxies in these
bins, which caused instabilities in the estimator, hence poor
convergence.  The red and blue bands display the distributions and
$68\%$ confidence ranges obtained by applying two other methods:
\begin{itemize}
\item Calibration with small-scale cross-correlations (CC), shown by
  the red bands and implemented by applying the recipes of
  \citet{2008ApJ...684...88N} and \citet{2010ApJ...721..456M} to the
  cross-clustering of KiDS sources and deep spectroscopic samples from
  DEEP2 and COSMOS;
\item Weighted direct calibration (DIR), shown by the blue bands and
  based on direct determination of the source redshift distribution by
  cross-matching with a number of overlapping spectroscopic samples,
  with appropriate re-weighting for incompleteness.
\end{itemize}
We refer the reader to \citet{2016arXiv160605338H} for more details
about these methods.  We converted the DIR estimation from
$P^{(p)}(z)$ to $b^{(p)}(z) \times P^{(p)}(z)$ using the function
$b^{(p)}(z)$ implicitly assumed in the CC method, as outlined by
\citet{2010ApJ...721..456M}.  For the purposes of this comparison, all
the distributions have been normalized such that $\int_{0.1}^{0.8} dz
\, b^{(p)} \, P^{(p)} = 1$.  These different methods produce redshift
distributions which agree in a qualitative sense, although comparisons
illustrate systematic errors affecting each technique.

The equivalent quadratic estimation for the KiDS-800 $i$-band imaging
data, analyzed in a single tomographic bin, is shown in
Fig. (\ref{iband_results}).  In this figure the blue points display
the reconstructed redshift distribution with 1-$\sigma$ errors, and
the shaded blue region shows the $95\%$ confidence interval for the GP
model.  We observe that the amplitude of the cross-correlation signal
increases with redshift, driven by a combination of the underlying
redshift distribution $P^{(p)}(z)$ and the source galaxy bias
$b^{(p)}(z)$.  Assuming that the latter is a slowly-varying function,
this analysis suggests that the redshift distribution of the $i$-band
sources is broad and peaked at $z \gtrsim 0.7$.  These findings are
qualitatively consistent with the DIR estimate presented by
\citetalias{amon2016}, although further comparison is beyond the scope
of the current study.

These results demonstrate that, although the 2dFLenS dataset does not
have sufficient redshift coverage to derive the full source
distributions by cross-correlation, we can successfully apply our
methodology to the KiDS dataset and obtain results in qualitative
agreement with previous determinations.

\section{Conclusion}
\label{sec:final}

With the issue of source redshift calibration becoming increasingly
pronounced for weak gravitational lensing surveys, new and versatile
approaches to this problem are needed.  Calibration via
cross-correlation with overlapping spectroscopic surveys represents
one such approach.  In this work, we have presented our efforts to
extend the accuracy and applicability of such methods to both
simulations and data.  We summarise our main findings as follows:

\begin{itemize}

\item{We have developed a new, minimum-variance and unbiased {\it
    quadratic estimator} that infers the redshift probability
  distributions of photometric samples of galaxies $P^{(p)}(z)$, in a
  degenerate combination with their galaxy bias $b^{(p)}(z)$, based on
  their angular cross-correlation with an overlapping spectroscopic
  sample.  This derivation expands on work presented by
  \citet{2013MNRAS.433.2857M}.}

\item{We have tested our methodology on a series of mock galaxy
  catalogues.  We found that at the level of statistical errors of
  current surveys the estimator is unbiased.  However, if we stack
  mocks together to perform a more accurate test, we discover small
  but significant discrepancies -- we attribute these effects to the
  breakdown of our clustering model due, for example, to non-linear
  galaxy bias.}

\item{We derive non-parametric, continuous functional forms of
  $b^{(p)} P^{(p)}(z)$ by building {\it Gaussian Process} models from
  the step-wise constraints inferred from the quadratic estimator.
  Such continuous functions are useful for modelling the lensing
  signal, and allow the computation of continuous confidence
  intervals.}

\item{We have applied our methodology to infer $b^{(p)} P^{(p)}(z)$
  functions for KiDS $r$-band and $i$-band imaging catalogues in the
  range $0.1 < z < 0.8$, through cross-correlation with the 2dFLenS
  spectroscopic redshift survey.  Our distributions are in qualitative
  agreement with the results of other methods.}

\end{itemize}

Our analysis could be extended in a number of ways: by improving the
modelling of non-linear effects, enhancing our mock catalogues to
match our datasets more closely, and re-formulating the estimator to
measure the spectroscopic galaxy bias and redshift distribution
simultaneously.  However, the calibration of the bias of the
photometric sample remains the most critical component.  Possible
approaches to this problem include the use of redshift-space
distortions, lensing magnification, and galaxy-galaxy lensing.  We
hope that our work motivates more research on these topics.

\section*{Acknowledgments}

We thank Matt McQuinn and Martin White for very helpful discussions.

Parts of this research were conducted by the Australian Research Council 
Centre of Excellence for All-Sky Astrophysics (CAASTRO), through project 
number CE110001020.

CB acknowledges the support of the Australian Research Council through
the award of a Future Fellowship.

JHD acknowledges support from the European Commission under a
Marie-Sk{\l}odwoska-Curie European Fellowship (EU project 656869) and
from the NSERC of Canada.

CH acknowledges funding from the European Research Council under grant
number 647112.

HH and CBM are supported by an Emmy Noether grant (No.\ Hi 1495/2-1)
of the Deutsche Forschungsgemeinschaft.

DK is supported by the Deutsche Forschungsgemeinschaft in the
framework of the TR33 `The Dark Universe'.

KK acknowledges support by the Alexander von Humboldt Foundation.

DP is supported by an Australian Research Council Future Fellowship
[grant number FT130101086].

The 2dFLenS survey is based on data acquired through the Australian
Astronomical Observatory, under program A/2014B/008.  It would not
have been possible without the dedicated work of the staff of the AAO
in the development and support of the 2dF-AAOmega system, and the
running of the AAT.

This study is based on data products from observations made with ESO
Telescopes at the La Silla Paranal Observatory under programme IDs
177.A-3016, 177.A-3017 and 177.A-3018, and on data products produced
by Target/OmegaCEN, INAF-OACN, INAF-OAPD and the KiDS production team,
on behalf of the KiDS consortium. OmegaCEN and the KiDS production
team acknowledge support by NOVA and NWO-M grants. Members of
INAF-OAPD and INAF-OACN also acknowledge the support from the
Department of Physics \& Astronomy of the University of Padova, and of
the Department of Physics of Univ.\ Federico II (Naples).

This work was performed on the gSTAR national facility at Swinburne
University of Technology.  gSTAR is funded by Swinburne and the
Australian Government’s Education Investment Fund.

Computations for the $N$-body simulations were performed in part on
the Orcinus supercomputer at the WestGrid HPC consortium
(www.westgrid.ca), in part on the GPC supercomputer at the SciNet HPC
Consortium. SciNet is funded by: the Canada Foundation for Innovation
under the auspices of Compute Canada; the Government of Ontario;
Ontario Research Fund - Research Excellence; and the University of
Toronto.

\end{document}